\documentclass[letterpaper]{article} % DO NOT CHANGE THIS
\usepackage[preprint]{aaai2027} 
\usepackage[hyphens]{url}  % DO NOT CHANGE THIS
\usepackage{graphicx} % DO NOT CHANGE THIS
\usepackage{natbib}  % DO NOT CHANGE THIS AND DO NOT ADD ANY OPTIONS TO IT
\usepackage{caption} % DO NOT CHANGE THIS AND DO NOT ADD ANY OPTIONS TO IT
\usepackage{algorithm}
\usepackage{newfloat}
\usepackage{listings}
\DeclareCaptionStyle{ruled}{labelfont=normalfont,labelsep=colon,strut=off} % DO NOT CHANGE THIS
\floatstyle{ruled}
\newfloat{listing}{tb}{lst}{}
\floatname{listing}{Listing}

\usepackage{booktabs}

\usepackage{amsmath}
\usepackage{amssymb}
\usepackage{booktabs}
\usepackage[table]{xcolor}  

\newcommand{\ie}{\emph{i.e., }}
\newcommand{\eg}{\emph{e.g., }}

\newcommand{\etc}{\emph{etc.}}

\usepackage{algpseudocode}
\usepackage{array}
\usepackage{enumitem}
\usepackage{subcaption}
\usepackage{caption}
\usepackage{tabularx}
\usepackage{colortbl}
\usepackage{makecell}
\usepackage{placeins}
\usepackage{titletoc}
\usepackage{arydshln}
\usepackage[most]{tcolorbox}

\definecolor{groupgray}{gray}{0.93}
\definecolor{lessonblue}{RGB}{235,242,250}
\definecolor{lessongray}{RGB}{245,245,245}
\definecolor{lessonborder}{RGB}{40,40,40}

\newtcolorbox{learningbox}[2]{
  enhanced,
  breakable,
  colback=white,
  colframe=lessonborder,
  boxrule=0.8pt,
  arc=1pt,
  left=6pt,
  right=6pt,
  top=6pt,
  bottom=6pt,
  title=\textbf{#1: #2},
  coltitle=black,
  colbacktitle=lessongray,
  fonttitle=\bfseries,
  attach boxed title to top center={yshift=-1mm},
  boxed title style={
    colback=lessongray,
    colframe=lessongray,
    boxrule=0pt,
    arc=0pt
  }
}

\newcommand{\AppendixTOC}{%
  \renewcommand{\addcontentsline}[3]{%
    \addtocontents{##1}{\protect\contentsline{##2}{##3}{\thepage}{}}}%
  \vspace{0.8em}
  \startcontents[appendix]%

  \noindent\makebox[\linewidth][c]{%
    \begin{minipage}{0.98\linewidth}
      {\large\bfseries Table of Contents}\par
      \vspace{0.6em}\noindent\hrule
      \titlecontents{section}
        [1.2em]{\addvspace{0.9em}}{\bfseries\thecontentslabel\quad}{}{\dotfill\contentspage}[]

      \titlecontents{subsection}
        [3.0em]{\addvspace{0.0em}}{\thecontentslabel\quad}{}{\dotfill\contentspage}[]

      \printcontents[appendix]{l}{1}{\setcounter{tocdepth}{2}}

      \vspace{0.8em}\noindent\hrule\par
    \end{minipage}%
  }

  \vspace{0.3em}
}

\title{Self-Evolving Multi-Agent Symbolic Discovery for Financial Fundamental Analysis}
\author{
    Kelvin J.L. Koa\textsuperscript{\rm 1, \rm 2}\corresponding,
    Filip Orestav\textsuperscript{\rm 3},
    Shengqiong Wu\textsuperscript{\rm 4}\corresponding,
    Michael J. Wooldridge\textsuperscript{\rm 4},
    Ke-Wei Huang\textsuperscript{\rm 1, \rm 2}
}

\affiliations{
    \textsuperscript{\rm 1}National University of Singapore
    \qquad
    \textsuperscript{\rm 2}Asian Institute of Digital Finance\\
    \textsuperscript{\rm 3}KTH Royal Institute of Technology
    \qquad
    \textsuperscript{\rm 4}University of Oxford\\
    \vspace{3px}
    kelvin.koa@u.nus.edu, orestavfilip@gmail.com, shengqionggwu@gmail.com\\
    michael.wooldridge@cs.ox.ac.uk, dishkw@nus.edu.sg
}

\begin{document}

\maketitle

\begin{abstract}
While symbolic regression (SR) has been successfully used in science to discover new equations, its use in financial valuation is hindered by several limitations. Whereas the natural sciences provide objectively correct relationships, financial valuation constitutes a distinct class of symbolic discovery problems, as it admits multiple valid perspectives, operates under non-stationary market conditions, and involves noisy, continuous performance signals. In this work, we propose \textbf{Mu}lti-Agent \textbf{F}undamental \textbf{A}nalysis with \textbf{S}ymbolic \textbf{A}daptive learning (\textsc{mufasa}), a hierarchical multi-agent framework for symbolic discovery in finance. \textsc{mufasa} introduces (1) disentangled equation discovery via specialized agents representing distinct valuation perspectives, (2) a meta-coordinator that performs hierarchical-level reasoning over market context information, and (3) a memory mechanism that reasons over statistical performance summaries (\eg accuracy, stability, and tail risk) to guide learning under noisy feedback. Experiments across datasets
from multiple countries show that \textsc{mufasa} achieves state-of-the-art performance on the valuation task compared to classical finance methods, financial large language models, and SR approaches, while simultaneously producing interpretable equations, which we share with the community. We also make publicly available the distilled learnings across evolution iterations and context-dependent strategy weights, which might offer useful insights for future research on financial fundamental analysis.
\end{abstract}

\section{Introduction} 
Fundamental analysis (FA) aims to estimate the intrinsic value of a company using accounting information such as earnings, cash flows, and assets, and plays a central role in investment decision-making \cite{bauman1996review}. A long-standing goal in this domain is to derive interpretable relationships (expressed as valuation formulas) that connect financial variables to future returns, which provides actionable insights for decision-making \cite{shmueli2010explain}. 
However, constructing such relationships remains a largely manual and iterative process, often relying on domain expertise to propose, test, and refine hypotheses.
Symbolic regression (SR) offers a promising direction for automating this process by discovering mathematical expressions directly from data \cite{cranmer2023interpretable}. 
Recent advances further integrate large language models (LLMs) into SR pipelines, enabling more efficient hypothesis generation and search by leveraging learned priors \cite{grayeli2024symbolic,shojaee2024llm,saveliev2026influence}. 
These methods have been particularly successful in scientific domains \cite{romera2024mathematical,novikov2025alphaevolve,xie2026language}, where the objective is to recover a single, stable governing equation from data.

However, unlike the natural sciences,  \textbf{financial valuation constitutes a fundamentally different class of symbolic discovery problems}, for which existing SR formulations are structurally misaligned. Specifically, symbolic discovery in the financial domain exhibits three defining characteristics:
% \vspace{-2px}
\begin{itemize}
\item \textbf{Multiple valid perspectives.} Unlike physical systems, financial valuation does not admit a single ground-truth equation. Different valuation philosophies (\eg earnings-based \cite{buffett2001essays}, cash flow-based \cite{damodaran2012investment}, growth-based \cite{asness1997parallels}) can simultaneously provide equally valid but distinct explanations of asset prices.
Current agent-based SR methods \cite{shojaee2024llm, romera2024mathematical} that search for a single equation may largely limit expressiveness and display reduced generalizability across different market conditions. 

\item \textbf{Non-stationary environments.} The relationship between financial signals and outcomes varies across market conditions, such as market regimes and company sectors. Because of this, the same actions could lead to different results across learning iterations. The reward pipeline is confounded by the unstable environment, making it difficult to reliably transfer or reuse past experience.

\item \textbf{Noisy, continuous feedback.} Candidate expressions are evaluated through noisy and continuous performance signals rather than binary correctness signals. 
% This complicates the learning of effective search strategies from past experience, as current agent memory systems \cite{ouyang2025reasoningbank, zhang2025memevolve} typically distill lessons from binary success and failure learning signals to inform agent decisions.
Current agent memory systems \cite{ouyang2025reasoningbank, zhang2025memevolve,allard2026experiential,xiong2025memory} that distill lessons from binary success and failure signals may misattribute success to spurious patterns arising from noise or distill learnings that are not robust across the noisy dataset.
\vspace{-2px}
\end{itemize}

\begin{figure*}[t]
\centering
\vspace{-20px}
\includegraphics[width=0.95\textwidth]{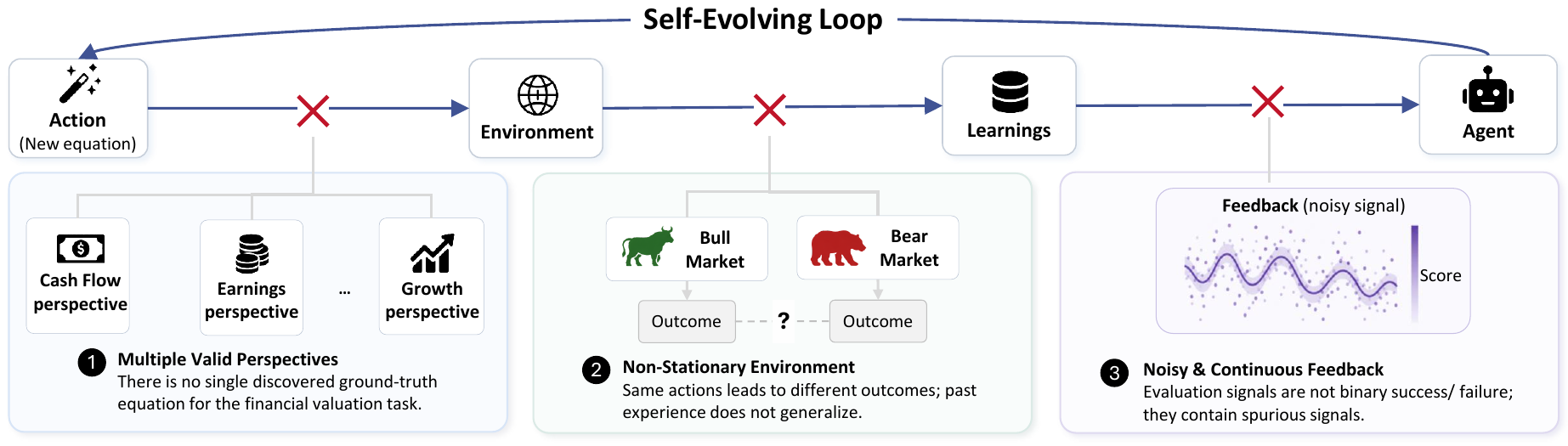}
\vspace{-3px}
\caption{Overview of the challenges in financial SR across different levels of the self-evolving loop.}
\label{fig:mufasa-again}
\vspace{-8px}
\end{figure*}

Importantly, these challenges also \textit{compound} across stages of the agent self-evolving loop (see Figure \ref{fig:mufasa-again}), making the overall symbolic FA progressively harder as a whole.

To address these challenges, we propose \textbf{Mu}lti-Agent \textbf{F}undamental \textbf{A}nalysis with \textbf{S}ymbolic \textbf{A}daptive learning (\textsc{mufasa}), a hierarchical multi-agent framework for symbolic discovery in finance. 
First, instead of searching for a single universal equation, \textsc{mufasa} performs \underline{disentangled} symbolic discovery via a set of specialized agents, each operating under a distinct valuation perspective (\eg cash flow, earnings, \etc) 
This design encourages diversity in the discovered symbolic relationships while maintaining interpretability within each perspective. 
Second, to account for non-stationary market conditions, we perform \underline{hierarchical-level} reasoning by introducing a meta-coordinator agent that conditions on contextual information (\eg sectors, regimes) to dynamically weigh the outputs of specialist agents. 
This enables an adaptive combination of perspective-specific knowledge across varying environments. 
Third, to cope with noisy and continuous feedback, each agent maintains a structured memory of previously explored equations and summaries of their \underline{statistical} performance. By reasoning over signals such as accuracy, stability, tail risk, \etc, agents can identify robust candidates and refine their search strategies despite noisy supervision from individual observations.

To demonstrate the effectiveness of \textsc{mufasa}, we perform extensive experiments across companies from multiple countries (\eg US, UK, China), using their fundamental information obtained from Bloomberg. 
Our model achieves state-of-the-art value forecasting performance compared to \textit{classic financial valuation methods}, \textit{financial LLMs}, and \textit{symbolic regression methods}, while also producing interpretable equations, which we share with the research community. In addition, we also make publicly available the distilled lessons across evolution iterations and the importance weights of each valuation method across regimes and sectors, which contain insights that could be useful for financial practitioners.  
In summary, our contributions in this work are:
\begin{itemize}
\item We identify financial valuation as a distinct class of symbolic regression problems, characterized by multi-perspective validity, non-stationary environments, and noisy, continuous feedback.

\item We introduce \textsc{mufasa}, a hierarchical framework that decomposes symbolic discovery across specialized agents, performs hierarchical-level reasoning over environmental signals, and distills lessons from statistical feedback rather than binary signals.

\item We show that \textsc{mufasa} achieves state-of-the-art valuation forecasting performance across multiple global markets, while recovering interpretable and economically grounded symbolic relationships.

\item We release a rich set of artifacts, including the discovered symbolic equations, distilled lessons across self-evolving iterations, and context-dependent importance of different valuation perspectives, offering insights for future research on financial fundamental analysis.
\end{itemize}
\section{Related Work}
\paragraph{Financial Fundamental Analysis.} Traditional finance literature defines fundamental analysis (FA) as the evaluation of a company’s accounting information \cite{bauman1996review}, such as the information found in financial statements or balance sheets. The aim is to understand the intrinsic value of the company \cite{graham1951security} and identify stocks that may be undervalued to make a profit. There are multiple approaches to fundamental investing, such as value investing \cite{graham2003intelligent, buffett2001essays}, factor investing \cite{fama1992cross, asness1997parallels}, or investing around macroeconomic trends \cite{dalio2018principles}. Each of these involves analyzing different types of financial information and may result in different investment outcomes \cite{capaul1993international, asness2000style, petkova2005value}. Unlike traditional deep-learning methods \cite{xu2018stock, hu2018listening, lin2021learning}, financial approaches emphasize economic interpretability \cite{shmueli2010explain}. Recent works have also explored using financial LLMs to perform interpretable analysis over structured accounting data \cite{qian2025fino1, liu2025fin}, but these focus on financial QA, and do not deal with forecasting or financial valuation.

\paragraph{Symbolic Regression.} Symbolic regression (SR) aims to discover mathematical expressions that describe relationships in data \cite{cranmer2023interpretable}. Classical SR approaches \cite{schmidt2009distilling, schmidt2009symbolic} are often based on evolutionary algorithms, such as genetic programming, which iteratively evolve candidate expressions through mutation and recombination. While effective, these methods often suffer from large search spaces and computational inefficiency. The emergence of large language models (LLMs) advanced SR by enabling knowledge-guided hypothesis generation, where models can propose equations informed by prior scientific knowledge \cite{grayeli2024symbolic, shojaee2024llm}. These approaches have demonstrated strong performance in rediscovering known physical laws and uncovering novel relationships in scientific domains \cite{romera2024mathematical, novikov2025alphaevolve,xie2026language}.

\paragraph{Self-Evolving LLM Agents.} Early works have shown that LLMs can function as agents that interact with external tools and environments \cite{yao2022react, schick2023toolformer}, which established the foundation for LLM agents, unlocking external capabilities beyond learning model weights. Later works show that LLM agents can self-improve through various techniques: For example, self-reflective frameworks enable agents to improve by learning from feedback signals from the environment \cite{shinn2023reflexion, zelikman2022star}. In contrast, agent memory enables agents to accumulate experience over time and refine reasoning and decision-making strategies \cite{zhang2025memevolve, cao2025remember, long2025seeing}. This idea of using LLMs as agents is further extended to self-evolving multi-agent systems, where multiple agents can collectively improve through coordination \cite{wang2025evoagentx, zhai2025agentevolver, xia2025agent0}. Within these systems, different agents can take on specialized roles, while the overall system would improve through environment feedback, experiential memory, and interaction across the agents \cite{qiu2025alita, qiu2025alitag, zhang2025agentorchestra}.

\begin{figure*}[!th]
\vspace{-20px}
\centering
\includegraphics[width=0.85\textwidth]{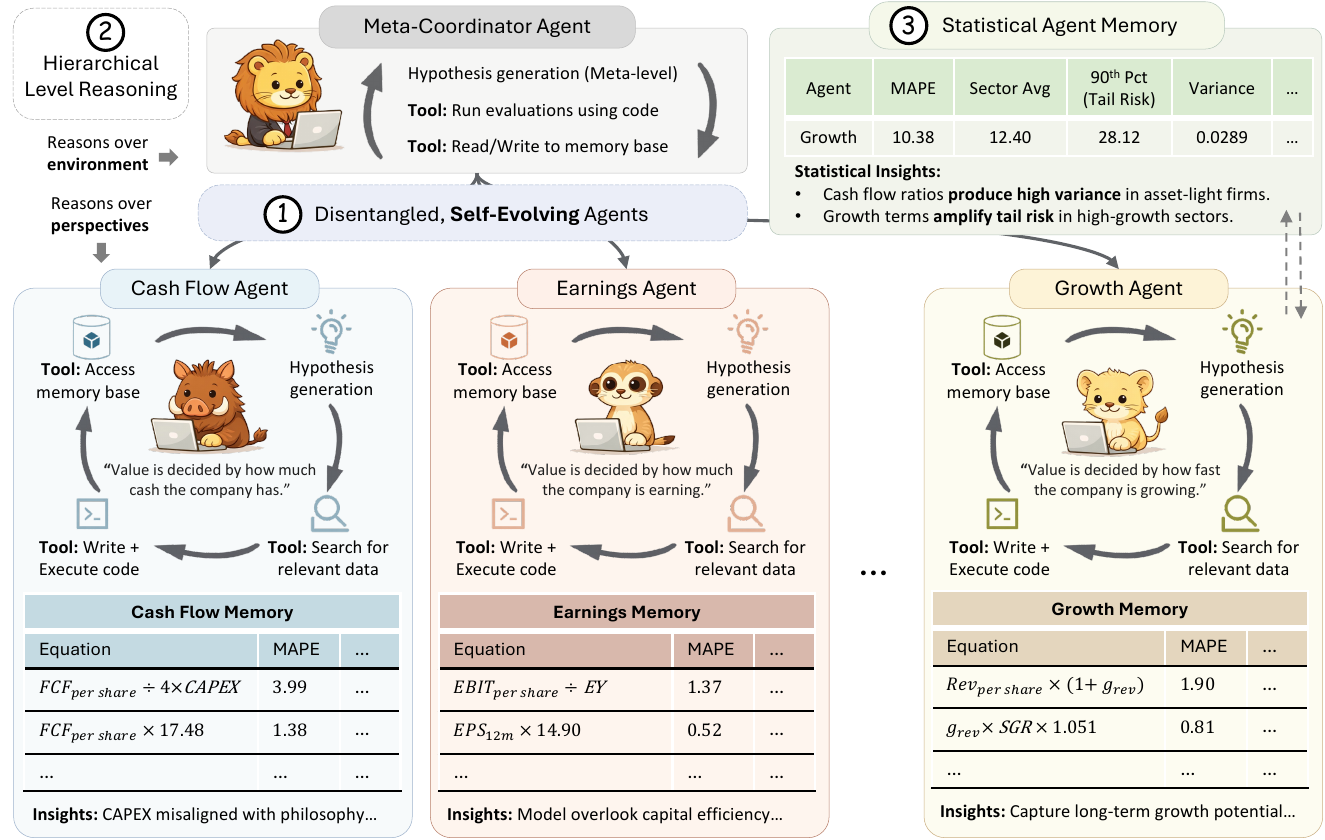}
\vspace{-4px}
\caption{Overview of the \textsc{mufasa} framework. \textsc{mufasa} uses disentangled multi-agent systems to encourage symbolic expression diversity. A meta-coordinator agent reasons over the non-stationary environment to integrate these expressions. Each agent maintains a memory to identify past statistical patterns.}
\label{fig:mufasa-framework}
\vspace{-8px}
\end{figure*}

\vspace{-2px}
\section{The \textsc{mufasa} Framework}
\vspace{-2px}
\label{sec:model}
In this section, we first study the interpretable fundamental analysis problem and its challenges as compared to the standard symbolic regression task. We then explain the proposed \textbf{Mu}lti-Agent \textbf{F}undamental \textbf{A}nalysis with \textbf{S}ymbolic \textbf{A}daptive learning (\textsc{mufasa}) framework, shown in Figure \ref{fig:mufasa-framework}.

\subsection{Problem Formulation}
Given a universe of companies and an accounting information database $\mathcal{X}_t$ for time $t$, our goal is to produce valuations for each company $i$, which are evaluated against their look-ahead stock prices $y_{i, t+1}$. 
Importantly, we aim to identify an \textit{interpretable} symbolic relationship, formalized by a symbolic expression $\tilde{f}$, that explains how company valuations are derived from accounting data.

This setting differs from the standard symbolic regression task \cite{cranmer2023interpretable, shojaee2024llm} in two ways. 
Firstly, the relevant explanatory variables $\mathbf{x}_{i, t} \subseteq \mathcal{X}_t$ are not explicitly specified a priori, but must instead be identified from the information database during the discovery process. 
Secondly, the underlying financial relationships are context-dependent, \ie the symbolic expression $\tilde{f}$ can vary across market conditions. 
We define a context variable $C(\cdot)$, which can incorporate multiple contextual factors. 
In this work, we define $C(i, t) = \left(s(i), r(t)\right)$, representing company sectors $s(i)$ and market regimes $r(t)$, respectively.
% We define two types of contexts: market regimes $r(t)$ and company sectors $s(i)$.
More formally, given $\mathcal{X}_t$ and observed $y_{i, t+1}$ for all companies $i$, our model seeks to discover a set of explanatory variables $\mathbf{\hat{x}}_{i, t}$ and the context-dependent symbolic expressions $\tilde{f}_{C(i,t)}$ such that:
\begin{equation}
\tilde{f}_{C(i,t)}(\mathbf{\hat{x}}_{i, t}) \approx y_{i,t+1}, \quad \forall i,t.
\end{equation}
The discovered variables and expressions should not only accurately fit the observed targets, but also maintain strong generalizability across unseen out-of-sample data while remaining interpretable.

\subsection{Disentangled Symbolic Discovery}

Unlike the natural sciences, financial valuation does not admit a single universal functional form.
% Instead, different economic mechanisms can give rise to distinct symbolic relationships between accounting variables and valuation targets.
% asness1997parallels, 
To capture this heterogeneity, we disentangle the symbolic discovery process using a set of perspectives, each associated with a specific valuation philosophy established in finance literature \cite{buffett2001essays, dalio2018principles}.
Then, each perspective is incorporated into an LLM agent, which seeks to find the best equation based on its beliefs. 
Each agent operates autonomously: it can propose hypotheses, 
retrieve relevant financial variables via external tools, and evaluate candidate expressions based on observed outcomes via code.

% This decomposition enables the exploration and discovery of multiple interpretable relationships without enforcing a single unified expression.

\paragraph{Perspective Decomposition.}
Formally, we define a set of valuation philosophies $\mathbf{a}$ that constrains the hypothesis space. 
Each philosophy induces inductive biases over candidate equations, including preferred variables and structural patterns, which prevents the collapse of expression diversity.

\paragraph{Hypothesis Generation.}
For each perspective $\mathbf{a}$, we employ an LLM $\pi_\theta$ to iteratively generate candidate symbolic expressions $\tilde{f}^{(\mathbf{a})}$. 
At each step, a new hypothesis is sampled as: $\tilde{f}^{(\mathbf{a})} \sim \pi_\theta(\cdot \mid \mathbf{p}^{(\mathbf{a})})$, where $\mathbf{p}^{(\mathbf{a})}$ is the constructed prompt. 
The prompt encodes the investing philosophy $\mathbf{a}$ and accumulated learnings $\mathbf{Z}^{(\mathbf{a})}$ derived from historical interactions and statistical feedback.
% generated from past statistical memory.

% To balance exploration and exploitation, the generation process is structured into three distinct phases: \textbf{Exploration}, which encourages diverse hypothesis proposals by reasoning from first principles; \textbf{Refinement}, which focuses on improving the current hypothesis by introducing new variables, adjusting parameters, or restructuring the equation; and \textbf{Fine-tuning}, which performs minor adjustments to further improve the best-performing equation. This procedure is applied across all training steps for each iteration, allowing the model to refine symbolic expressions based on aggregated feedback.

\paragraph{Tool Use: Data Retrieval.}
Each candidate expression $\tilde{f}^{(\mathbf{a})}$ implicitly defines a set of variables $\mathbf{\hat{x}}^{(\mathbf{a})}_{i,t}$ that the agent may retrieve from the database $\mathcal{X}_t$ using external tools. 
Specifically, the agent first queries a semantic retrieval mechanism to obtain candidate variable names and descriptions, and subsequently selects and retrieves the corresponding data. If a retrieval attempt fails (\eg due to unavailability or insufficient coverage), the model receives structured feedback \cite{schick2023toolformer} and adapts its hypothesis accordingly, either by refining the variables used or modifying the proposed expression.

\paragraph{Tool Use: Code Evaluation.}
Each candidate expression is represented as an executable program, enabling direct evaluation. Given the hypothesized expression $\tilde{f}^{(\mathbf{a})}$ and retrieved variables $\mathbf{x}^{(\mathbf{a})}_{i,t}$, the expression is evaluated to compute predicted values:
\begin{equation}
\hat{y}^{(\mathbf{a})}_{i,t+1} = \tilde{f}^{(\mathbf{a})}\left(\mathbf{x}^{(\mathbf{a})}_{i,t}\right).
\end{equation}
Similarly, if the expression fails to execute (\eg due to syntax errors, invalid operations, or non-finite outputs), the system returns structured feedback \cite{gao2023pal}, which is incorporated into the prompt $\mathbf{p}^{(\mathbf{a})}$. The agent then revises its subsequent hypothesis accordingly.

\paragraph{Hypothesis Optimization and Assessment.}
Similar to LLM-SR \cite{shojaee2024llm}, we refine candidate expressions through numerical optimization of their coefficients. Given a proposed expression $\tilde{f}^{(\mathbf{a})}$, we treat any scalar multipliers or parameters as continuous variables and optimize them directly with respect to the training data. This decouples structure discovery from parameter fitting: the LLM defines the functional form of $\tilde{f}^{(\mathbf{a})}$, while numerical routines adjust its coefficients to improve empirical performance. We use differential evolution \cite{storn1997differential} as our optimization approach.

Following optimization, we assess the fitness of each candidate expression by evaluating its predictive performance on observed data. Given the predicted target values $\hat{y}^{(\mathbf{a})}_{i,t+1}$, we define a fitness evaluation score as the prediction error against ground truth:
\begin{equation}
\mathcal{L}^{(\mathbf{a})} = \text{MAPE}\left(\hat{y}^{(\mathbf{a})}_{i,t+1}, y_{i,t+1}\right).
\end{equation}
Here, the output of the philosophy-constrained function is evaluated against the overall look-ahead prices $y_{i,t+1}$ to ensure that the most generalizable function for each philosophy is obtained, as detailed in Algorithm~1 of Appendix~\ref{detailed_methods}.

% \label{Algorithm_1}

\subsection{Hierarchical-Level Reasoning}

Financial market environments are inherently non-stationary. Because of this, the relationship between candidate solutions and their evaluation outcomes is unstable, making it difficult for learning experience to accumulate. To account for this, we introduce hierarchical-level reasoning, where a meta-coordinator agent learns to reason over market conditions to contextualize the different agent outputs. 

% In this work, we consider two types of contexts: $s(i) \in \mathcal{S}$ denotes the sector of company $i$, and $r(t) \in \mathcal{R}$ denotes the market regime at time $t$. The regime function $r(t)$ is obtained via a predefined mapping over market indicators, while $s(i)$ is given by the sector classification of each company. The design of the framework makes it fully generalizable across more types of contexts in future work.

For each context $C(i,t)$, the goal of the meta-coordinator agent $\pi_\phi$ is to produce a weight vector that shows the relative importance of the agents during that context. At each step, the meta-coordinator constructs a prompt $\mathbf{p}^\text{meta}$ which includes the current context, the agent-level performances $\mathcal{L}^{(\mathbf{a})}$  and the accumulated learnings $\mathbf{Z}^{\text{meta}}$. 
% It then produces the candidate weight vector: $\mathbf{\tilde{w}}_{C(i,t)} \sim \pi_\phi(\cdot \mid \mathbf{p}^\text{meta})$.
It then selects a subset of agents, and computes their weight vector $\mathbf{\tilde{w}}_{C(i,t)}$ via SLSQP-based \cite{kraft1988software} constrained optimization.

Given the agent outputs and the weight vector, we define the aggregated valuation in each context as:
\begin{equation}
\hat{y}_{i,t+1} = \mathbf{\tilde{w}}_{C(i,t)}^\top \cdot\hat{\mathbf{y}}_{i,t+1},
\end{equation}
% where $\hat{\mathbf{y}}_{i,t+1}$ represents the output vector of outputs from all philosophy-disentangled agents.
where $\hat{\mathbf{y}}_{i,t+1}$ denotes the vector collecting $\hat{y}^{(\mathbf{a})}_{i,t+1}$ for the set of philosophy-disentangled agents $\mathbf{a}$.

Similarly, we assess the quality of each candidate weight vector by evaluating its predictive error:
\begin{equation}
\mathcal{L}^{\text{meta}} = \text{MAPE}\left(\hat{y}_{i,t+1}, y_{i,t+1}\right).
\end{equation}
This hierarchical design enables continual adaptation of agent weights under non-stationary feedback.

\subsection{Statistical Performance Memory}

Unlike the natural sciences, which seek a single correct expression, or agent-based tasks with binary outcomes, financial expressions are evaluated through noisy and continuous performance measures. This complicates the ability of current agent memory \cite{ouyang2025reasoningbank, zhang2025memevolve} to effectively learn from past experience, as they typically rely on signals derived from successful or failed attempts at completing a task.
To cope with this, we augment agent memory with \textit{structured statistical information} that captures the statistical behavior of previously explored equations. Rather than storing only a scalar loss, memory records each hypothesis together with a statistical summary of its observed behavior, as demonstrated in Table~\ref{tab:memory_stats}. 
\begin{table*}[!th]
\vspace{-4px}
\begin{center}
\footnotesize
\resizebox{0.99\textwidth}{!}{
\begin{tabular}{m{0.14\textwidth} m{0.40\textwidth} m{0.42\textwidth}}
\toprule
\textbf{Category} & \textbf{What is stored} & \textbf{Why} \\[-0.4em]
\midrule

Accuracy & Mean MAPE, Median MAPE & Captures overall predictive accuracy while reducing sensitivity to outliers and skewed observations. \\
\addlinespace[0.4em]

Stability & Standard Deviation, IQR, Coefficient of Variation & Measures whether performance is consistent or highly variable across samples or time periods. \\
\addlinespace[0.4em]

Tail Risk & $p90/p95$ Error, Worst-Decile Mean & Exposes extreme failure modes that are often hidden by average performance metrics. \\
\addlinespace[0.4em]

Complexity & Variable + Operator Count, Prediction Coverage & Reflects interpretability and numerical reliability of the expression in practical use. \\
\addlinespace[0.4em]

Bias & Mean Signed Percentage Error (MSPE) & Detects systematic over- or under-prediction across observations rather than isolated errors. \\
\addlinespace[0.4em]

Ranking & Spearman Rank Correlation & Captures whether the expression preserves cross-sectional ordering beyond absolute scale accuracy. \\
\addlinespace[0.4em]

Calibration & Signed Pct Error by Quartiles of Actual Price & Reveals whether the expression systematically misprices cheap or expensive stocks. \\

\bottomrule
\end{tabular}
}
\caption{Statistical information collected by \textsc{mufasa} to enable learning under continuous feedback.}
\label{tab:memory_stats}
\end{center}
\vspace{-4px}
\end{table*}

From the memory, we distill a set of learnings $\mathbf{Z}^{(\mathbf{a})}$. 
which guide the generation of new hypotheses at the beginning of the loop.
These learnings capture recurring statistical regularities that are not apparent from individual binary pass or fail signals, where a single evaluation could be dominated by noise and lead to unreliable scoring of candidate expressions. By aggregating outcomes across repeated evaluations, the system can recover more reliable estimates of relative performance, including properties such as variance and tail behavior that are not observable from isolated observations. 
% Some examples include:
% \begin{itemize}[leftmargin=*]
% \item \emph{Ratio-based expressions consistently exhibit lower result variance than additive forms.}
% \item \emph{Simpler expressions tend to avoid extreme tail errors compared to complex, composite ones.}
% % \item \emph{Earnings signals consistently perform better during bearish regimes than growth signals.}
% \end{itemize}
% These learnings are used to guide the generation of new hypotheses, going back to the start of the equation search process. 

Through this process (detailed in Algorithm~2 of Appendix~\ref{detailed_methods}), \textsc{mufasa} infers heuristically effective strategies that demonstrate consistent performance across its evaluations, improving the overall robustness of the symbolic search.
\section{Experiments} 
\label{sec:experiments}
\paragraph{Dataset.} 
\label{para:dataset}
We evaluate \textsc{mufasa} across five equity markets. 
The primary dataset
consists of all S\&P 500 constituents from 1990 Q1 to 2023 Q2, with constituent entry and exit dates included to prevent survivorship bias. To demonstrate generalisability across geographies and market structures, we additionally evaluate on the CSI 300 (2002-2021), STOXX 600 (1995-2025), Russell 2000 (1990-2025), and FTSE Asia Pacific ex-Japan (1996-2025), beginning each series at the earliest date with sufficient data 
coverage.
For all markets, data is split chronologically: the first 80\% of quarters form the training set and the remaining 20\% form the held-out test
set. 
The fundamental data is sourced from Bloomberg and the macroeconomic data primarily from FactSet. We use data from original quarterly filings only, preventing look-ahead bias from subsequent restatements, and define the
target variable as the stock price at the end of the fiscal quarter immediately following each filing period. The forecasting resolution is quarterly, as fundamental analysis is typically used for long-term stock investment.
\cite{graham2003intelligent}.

% \vspace{-14px}
\paragraph{Market Settings.} For regime context, bull and bear markets are defined at the index level following the standard practitioner convention \citep{pagan2003simple}: a bull market is declared when the index has risen at least 20\% from its most recent trough, and a bear market when it has fallen at least 20\% from its most recent peak. Regimes are computed from the index prices sourced from Yahoo Finance.

\vspace{-5px}
\paragraph{Baselines.} We compare \textsc{mufasa} against three categories of closely-related benchmark models. 
\begin{enumerate}

\item \textit{Traditional machine learning methods}: Random Forest~\cite{breiman2001random}, HistGBM~\cite{ke2017lightgbm}, and NeuralNet~\cite{rumelhart1986learning}. 

\item \textit{Symbolic regression}: PySR \cite{cranmer2023interpretable}, FunSearch \cite{romera2024mathematical}, and LLM-SR \cite{shojaee2024llm}. These models do symbolic discovery for scientific equations using evolutionary algorithms or an iterative single-LLM loop. 

\item \textit{Financial reasoning LLMs}: Fin-R1 \cite{liu2025fin} and Fino1 \cite{qian2025fino1}. These pre-trained LLMs were originally finetuned to perform reasoning over financial accounting data, but they do not deal with forecasting. 

\item \textit{Financial valuation}: We compare with some classical financial valuation approaches, including the sector-median P/E, EV/EBIT, EV/EBITDA, and P/B ratios, and also the Dividend Discount Model (DDM), Free Cash Flow to Equity (FCFE), and Free Cash Flow to Firm (FCFF) models \cite{damodaran2012investment}. 
% All baselines use the same train/test split and have access to the same feature set.

\item \textit{Financial agentic frameworks}: FinCon~\cite{yu2024fincon}; FinVision~\cite{fatemi2024finvision}
\end{enumerate}

\vspace{-5px}
\paragraph{Implementation Details.} \label{para:implementation}
All LLMs are served locally using vLLM on NVIDIA RTX A5000 GPUs, which takes \textasciitilde10 mins per agent on each processing unit.
We use Qwen3-4B-Thinking~\cite{qwen3technicalreport} for all agents, with a temperature of 0.7 for the agents and 0.3 for the memory components. The model context window is 32,768 tokens; we set a maximum generation length of 16,384 tokens.
We evaluate the models on \textbf{Mean Absolute Percentage Error (MAPE)}. The lower the results, the better.
\vspace{-5px}
\section{Results}

\begin{table*}[h]\footnotesize
\vspace{-15px}
% \resizebox{\textwidth}{!}{
\begin{tabular*}{\textwidth}{@{\extracolsep{\fill}}l
w{c}{1.15cm}w{c}{1.15cm}w{c}{1.15cm}w{c}{1.15cm}w{c}{1.15cm}w{c}{1.15cm}w{c}{1.15cm}}
\toprule
\textbf{Model} & \textbf{S\&P 500} & \textbf{Russell 2000} & \textbf{STOXX 600} & \textbf{FTSE APAC} & \textbf{CSI 300} & \textbf{All Data} & \textbf{Avg $\Delta$} \\ 
\midrule
\rowcolor{groupgray}\multicolumn{8}{c}{\textit{\textbf{Machine Learning Based Methods	}}} \\				
Random Forest &	0.4812 &	0.6551 &	0.7490 &	0.8436 &	0.5605 & 0.6579 & 0.1498 \\
HistGBM &	0.5916 &	0.8121 &	0.9774 &	1.3771 &	0.6041 & 0.8725 & 0.3644 \\
NeuralNet &	0.6130 &	1.0302 &	0.8963 &	1.5292 & 0.5242 & 0.9186 & 0.4105 \\
\rowcolor{groupgray}\multicolumn{8}{c}{\textit{\textbf{Financial LLMs}}} \\
Fino1 \cite{qian2025fino1}                         & 1.0230 & 2.1244 & 1.0958 & 1.2587 & 1.3388 & 1.3681 & 0.8600 \\
Fin-R1 \cite{liu2025fin}                        & 0.5263 & 0.7324 & 0.7437 & 0.8862 & 0.8095 & 0.7396 & 0.2315 \\
\rowcolor{groupgray}\multicolumn{8}{c}{\textit{\textbf{Financial Valuation}}} \\
P/E                            & 0.5127 & 0.8450 & 0.6209 & \underline{0.7439} & 0.6035 & 0.6652 & 0.1571 \\
EV/EBIT                        & 0.5580 & 0.9907 & \underline{0.6099} & 0.9150 & 0.6698 & 0.7487 & 0.2406 \\
EV/EBITDA                      & 0.5558 & 1.1589 & 0.6808 & 0.9314 & 0.6472 & 0.7948 & 0.2867 \\
P/B                            & 0.7177 & 0.8727 & 2.3048 & 0.8371 & 0.8401 & 1.1145 & 0.6064 \\
DDM                            & 0.7991 & 0.9370 & 1.0324 & 0.8691 & 0.6358 & 0.8547 & 0.3466 \\
FCFE                           & 0.9865 & 1.1827 & 1.3975 & 0.8700 & 0.8196 & 1.0513 & 0.5432 \\
FCFF                           & 1.2434 & 1.4313 & 1.5155 & 0.9097 & 0.8412 & 1.1882 & 0.6801 \\
\rowcolor{groupgray}\multicolumn{8}{c}{\textit{\textbf{Symbolic Regression}}} \\
PySR \cite{cranmer2023interpretable}                           & 0.6394 & \underline{0.5892} & 0.6741 & 0.7514 & \underline{0.5215} & \underline{0.6351} & 0.1270 \\
FunSearch \cite{romera2024mathematical}                     & 0.7146 & 1.1524 & 0.7693 & 0.7855 & 0.5735 & 0.7991 & 0.2910 \\
LLM-SR \cite{shojaee2024llm}                        & \underline{0.4761} & 0.7822 & 0.8269 & 0.9534 & 0.5967 & 0.7271 & 0.2190 \\
\rowcolor{groupgray}\multicolumn{8}{c}{\textit{\textbf{Financial Agentic Frameworks}}} \\					
FinCon~\cite{yu2024fincon} &	0.9581 &	0.9746 &	1.1822 &	0.6358 &	0.5784 & 0.8658 & 0.3577 \\
FinVision~\cite{fatemi2024finvision} &	0.8203	 & 0.5842	 & 1.2521	 & 0.9974	 & 0.9956 & 0.9299 & 0.4218 \\
\hline
\textsc{mufasa} (Ours)       & \textbf{0.4465} & \textbf{0.5718} & \textbf{0.5064} & \textbf{0.5814} & \textbf{0.4343} & \textbf{0.5081} & 0.3702 (Avg) \\
\bottomrule
\end{tabular*}
\caption{Performance comparison. The best baselines are underlined, and the best results are bolded.}
\label{tab:results}
\end{table*}

\paragraph{Performance Comparison.}
Table \ref{tab:results} reports the forecasting performance. We observe the following:
% ,itemsep=-0.1pt
\begin{itemize}[leftmargin=*]

\item  Fin-R1 consistently outperforms Fino1, likely because its emphasis on generalization and transfer better matches the forecasting setting. Nevertheless, both LLMs underperform most specialized baselines.

\item Traditional valuation ratios (\eg P/E, EV/EBIT) competitive across most markets. Their simple structure makes them relatively robust to estimation error, although their performance varies considerably across regions, suggesting that markets rely on different fundamental signals.

% These are simpler ratio-based valuations, which makes them less sensitive to estimation errors and more generalizable across different market conditions. However, their performance also varies substantially across markets. This shows that no single classical valuation method is uniformly reliable across regions, and that different markets may emphasize different fundamental signals.

\item More complex discounted cash-flow methods, including FCFF and FCFE, perform less well because they depend on noisy estimates such as growth rates and discount factors. We use the single-stage Gordon Growth Model \cite{gordon1962savings}, which assumes constant growth; more accurate parameter estimation may improve these methods.

% In contrast, more complex discounted cash flow variants such as FCFF and FCFE show weaker performance, as they rely on noisy estimated values (\eg growth rates and discount factors). For these baselines, we use a single-stage Gordon Growth Model \cite{gordon1962savings}, which assumes constant growth. These might improve with more accurate parameter estimations, which we leave for future work.

\item Among the SR baselines, PySR performs strongly, unlike the typical advantage of LLM-based SR. Scientific priors encoded by LLMs may be less useful in finance, where relationships are context-dependent and lack stable governing laws. PySR's evolutionary search may therefore be better suited to identifying empirically effective expressions.

% Among the SR baselines, PySR shows strong performance, which runs contrary to standard results. Typically, the LLM-based SR methods work by leveraging the scientific knowledge of the LLMs to propose better hypotheses. This strategy might work less effectively in the financial domain, where the relationships are more context-dependent and less governed by stable laws, reducing the usefulness of such priors. In contrast, PySR's evolutionary algorithms would work better in these cases by finding the equations that perform the best heuristically.

% \item Across all methods, performance varies noticeably between datasets. Errors are generally lower on S\&P 500 and CSI 300, and higher on FTSE APAC and Russell 2000. This may reflect differences in market structure, firm characteristics, and accounting regimes, as well as the increased noise typically observed in smaller-cap universes. These variations highlight the context-dependent nature of financial relationships, where a single global model may struggle to perform consistently across regions.

\item \textsc{mufasa} achieves the best performance across all datasets, with an average improvement of 19.65\%. Existing financial agentic frameworks, FinCon and FinVision, perform competitively in selected markets but generalize inconsistently across regions. In contrast, \textsc{mufasa} outperforms the better agentic baseline in every market, reducing its error by 27.67\% on average. This advantage stems from perspective-specific hypothesis generation and statistical memory that retains empirically robust expressions.

% The disentanglement of valuation perspectives re-allows the LLM agents to leverage their internal knowledge to propose better hypotheses based on existing theories on each perspective, while their statistical memory allows them to search for expressions that can perform well heuristically.

\end{itemize}
More details on the statistical significance of the experiments can be found in Appendix \ref{app:significance}.

\begin{table}[!th]
\small
% \vspace{-15px}
\centering
\resizebox{\linewidth}{!}{
\begin{tabular}{lccccc}
\toprule
\textbf{Model Variant} & \textbf{S\&P 500} & \textbf{Russell 2000} & \textbf{STOXX 600} & \textbf{FTSE APAC} & \textbf{CSI 300 }\\
\midrule
S.P Agent 
& 0.4971 & 0.8994 & 0.7608 & 0.6095 & 0.5928 \\

M.P (Equal Weights) 
& \underline{0.4652} & 0.8768 & 0.5448 & 0.6442 & 0.5309 \\

M.P (No Stats Memory) 
& 0.4702 & \underline{0.7525} & 0.5398 & 0.5852 & 0.4852 \\

M.P (Best Single Agent) 
& 0.4676 & 1.9173 & \underline{0.5256} & \underline{0.5833} & \underline{0.4739} \\

\textsc{mufasa} (Ours)
& \textbf{0.4465} & \textbf{0.5718} & \textbf{0.5064} & \textbf{0.5814} & \textbf{0.4343} \\

\bottomrule
\end{tabular}
}
\caption{Ablation study on perspective variant. S.P and M.P denote the single-/multiple perspective, respectively. Best results are bolded, second best are underlined.}
\label{tab:ablation}
\vspace{-7px}
\end{table}

\begin{table*}[!th]
\vspace{-9px}
\centering
\resizebox{\linewidth}{!}{%
\begin{tabular}{llll}
\toprule
\textbf{Model} & \textbf{Equation} & \textbf{Theory Equation} & \textbf{Theory} \\
\midrule

Fino1 &
$\dfrac{EV/EBITDA\cdot(1+g_{EPS}/100)}{EV/EBITDA_{mkt}\cdot EBITDA}$ &
- &
- \\[0.4em]

Fin-R1 &
$EPS_{TTM}\cdot PE$ &
$EPS\cdot PE$ &
Earnings Multiple \\[0.4em]

PySR &
$\sqrt{1.79\cdot |BVPS|\cdot\sqrt{\sqrt{|Tax|+Tax}\cdot\left|EPS\cdot e^{2.43}-e^{\sqrt{Y_{30}}}+\left(e^{PB_{mkt,med}}+EPS\right)\cdot4.23\right|}}$ &
- &
- \\[0.4em]

FunSearch &
$0.40\cdot EBITDA_{ps}+0.30\cdot Revenue_{ps}+0.20\cdot ROE+0.10\cdot PE$ &
- &
- \\[0.4em]

LLM-SR &
$0.01\cdot PE-0.03\cdot \dfrac{EV}{EBIT}-0.04\cdot \dfrac{EV}{EBITDA}+0.58\cdot PB+13.99\cdot NI_{ps}-0.01\cdot Revenue_{ps}$ &
$R_i=\alpha+\beta_1X_{1,i}+\beta_2X_{2,i}+\cdots$ &
Linear Factor \\[0.4em]

\cdashline{1-4}
\rowcolor{groupgray}\multicolumn{4}{l}{\textsc{mufasa} agents} \\
Earnings &
$NI_{ps}\cdot\left(1+0.01\cdot g_{EPS}\right)\cdot PE\cdot0.88$ &
$\dfrac{EPS}{r-g}$ &
Gordon Growth \\[0.4em]

Cashflow &
$FCF_{ps}\cdot \dfrac{EV}{EBITDA}\cdot\dfrac{100}{88.38}$ &
$\sum_{t=1}^{\infty}\dfrac{FCF_t}{(1+r)^t}$ &
Discounted Cash Flow \\[0.4em]

Asset &
$BVPS\cdot\left(1+1.01\cdot\dfrac{R\&D}{Assets}\right)\cdot\left(1-13829.10\cdot\dfrac{R\&D}{Assets}\right)\cdot\dfrac{PB-7.24\times10^{-4}}{1.34}$ &
$BVPS\cdot PB$ &
Book Multiple \\[0.4em]

Growth &
$(P/FCF)_{mkt}\cdot\left(FCF_{ps}+0.29\cdot\dfrac{P/FCF}{(P/FCF)_{mkt}}\right)\cdot\left(1-14.80\cdot(FCF_{ps}<1.08)\right)\cdot0.01$ &
$Metric\cdot Multiple$ &
Relative Valuation \\[0.4em]

Quality &
$NI_{ps}\cdot P/FCF\cdot\left(1+\log(1+ROE/100)\cdot0.87\right)$ &
$BV+\sum_{t=1}^{\infty}\dfrac{(ROE_t-r)\cdot BV_{t-1}}{(1+r)^t}$ &
Residual Income \\

\bottomrule
\end{tabular}%
}
\caption{Learnt equations \textit{vs.} related valuation forms in finance literature, for the S\&P500 dataset.}
\label{tab:learnt_equations}
\vspace{-8px}
\end{table*}

\vspace{-3px}
\paragraph{Ablation Study.} 
We conduct an ablation study to demonstrate the effectiveness of the model design.
\begin{itemize}
\vspace{-5px}
\item The \textsl{Single-Perspective Agent} uses one LLM agent to search for a unified valuation expression. It is outperformed by the equal-weight multi-perspective variant on four of the five markets, suggesting that representing heterogeneous valuation signals within a single search space limits effective exploration. 

% represents using only a single LLM agent to search for the best expression, which generally performed worse than the multi-perspective variants. Using one agent to search for equations in the financial domain is insufficient, as it forces heterogeneous valuation signals into a single search space and is difficult to optimize in dynamic market environments.

\item \textsl{Multi-Perspective (Equal Weights)} progresses to multi-perspective agents, but without hierarchical reasoning. It improves over the single-perspective variant. Disentangling into specialized agents helps optimization by focusing each agent on a subset of variables and relationships, enabling more diverse exploration and better generalization even when combined using simple averaging.

\item \textsl{Multi-Perspective (No Stats Memory)} adds hierarchical reasoning with the meta-coordinator agent, but without statistical memory. It outperforms equal weighting on four of the five markets, demonstrating the benefit of hierarchical coordination. Nevertheless, without statistical memory, the system relies primarily on coarse outcome feedback and cannot reliably distinguish consistently effective expressions from those that succeed by chance.
% It generally performs better than the equal-weight variant, showing that hierarchical reasoning is important for adaptively weighing different perspectives. However, binary feedback is limited, as it only indicates if an equation performs well, but not its consistency. This makes it harder to distinguish robust expressions from those that perform well by chance.

\item \textsl{Multi-Perspective (Best Single Agent)} retains the specialized agents and statistical memory but uses only the strongest individual agent. It obtains the second-best result on three markets but degrades substantially on Russell 2000, showing that reliance on a single perspective is unstable and requires an agent-selection rule that is not available in advance.

% utilizes multiple perspectives and statistical memory, but reports only the best single agent performance from the pool of disentangled agents. It achieves the second-best performance on most datasets. This shows that different perspectives perform well in different contexts, and the best agent proxies the performance of our individual agents. However, in practice, it is not possible to know beforehand which perspective would be best in each setting.

\item The \textsc{mufasa} framework achieves the best performance across all datasets. We see that integrating multiple perspectives shows improvement over any of the best agents. Overall, combining multi-perspective agents, hierarchical-level reasoning, and statistical memory produces the best results.
\end{itemize}
\vspace{-3px}
A \underline{more comprehensive} ablation study and additional experiments can be found in Appendix \ref{app:extended_ablation}, \ref{app:additional_results}.
% add some illustrations to show multi-perspective agents not coverage into one. 

\begin{figure}[!th]
    % \vspace{-5px}
    \centering
    % \resizebox{0.85\columnwidth}{!}{
        \begin{minipage}{\columnwidth}
            \centering
            % Row 1
            \includegraphics[width=0.32\columnwidth]{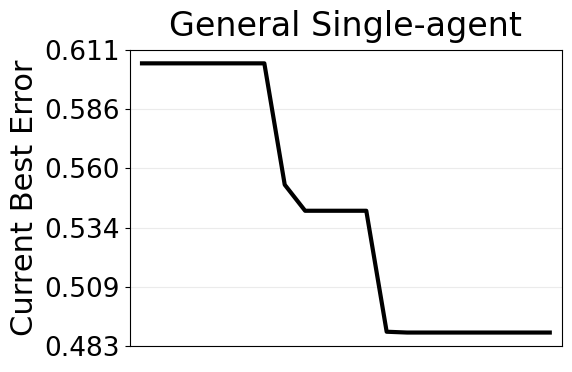}
            \includegraphics[width=0.32\columnwidth]{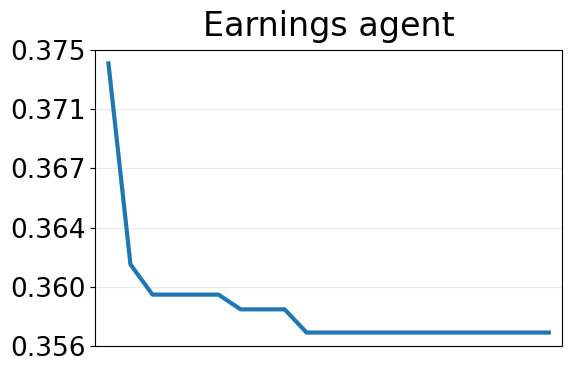}
            \includegraphics[width=0.32\columnwidth]{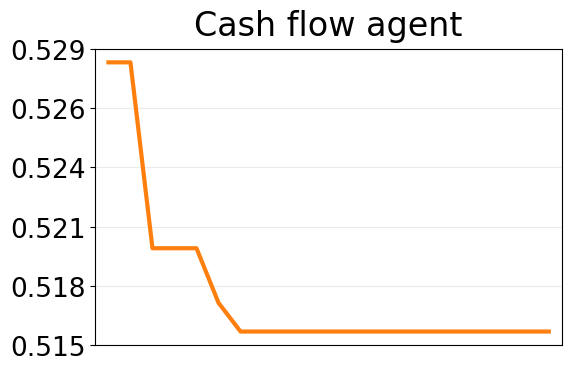}
            % Row 2
            \includegraphics[width=0.32\columnwidth]{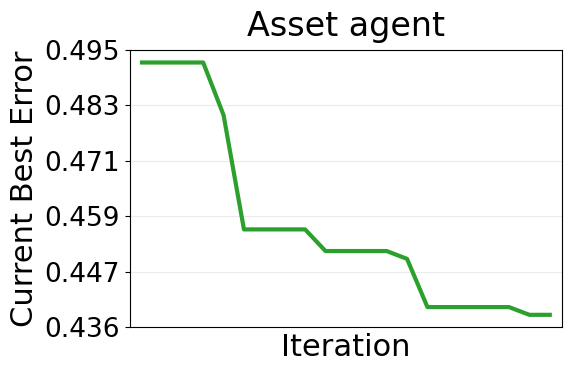}
            \includegraphics[width=0.32\columnwidth]{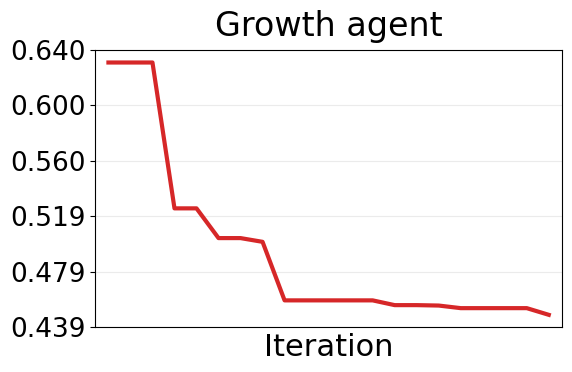}            \includegraphics[width=0.32\columnwidth]{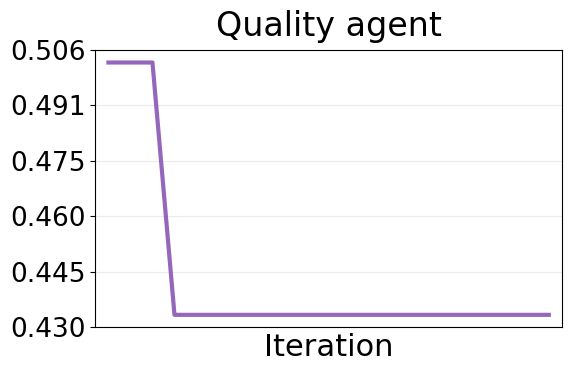}
        \end{minipage}
    % }
    % \vspace{-5px}
    \caption{Best error trajectories over the self-evolving iterations for each perspective agent.}
    % \vspace{-4px}
    \label{fig:agent_rolling_min}
\end{figure}

% \vspace{-10px}
\paragraph{Self-Evolving Agents.}
Figure \ref{fig:agent_rolling_min} shows the best error trajectories over iterations for each agent. Each agent exhibits a decrease in the lowest error over time, indicating that the symbolic search process iteratively refines candidate equations. Although the general single agent also improves, its final best error remains higher than those achieved by the specialized agents. This result is consistent with the difficulty of searching for a single expression that simultaneously captures heterogeneous valuation mechanisms. By restricting each agent to a specific valuation perspective, the disentangled design enables more focused exploration of the symbolic search space and yields stronger final solutions.

\paragraph{Discovered Equations.} Table \ref{tab:learnt_equations} shows the learnt equations on the S\&P 500 dataset, together with the most closely related valuation forms that can be found in finance literature \cite{damodaran2012investment, graham1951security}. We observe: 

\begin{itemize}[leftmargin=*,itemsep=-0.0pt]
\item Fin-R1 largely reproduces the standard earnings-multiple formula $EPS \cdot PE$. While interpretable, this fixed form offers limited adaptation to the observed data and represents only one valuation perspective.

% Fin-R1 directly uses the known earnings multiple formula $EPS \cdot PE$, as it was prompted in a zero-shot manner to produce interpretable forecasts. This limits its ability to adapt to different signals in the training data and also restricts it to a fixed valuation perspective, reducing its effectiveness.

\item PySR discovers a high-performing but financially opaque expression, reflecting its primarily heuristic search. LLM-SR instead combines variables from several perspectives in a linear form, but does not recover the perspective-specific structures observed in classical valuation models.

% PySR previously achieved second-best average performance on the main results, but its discovered expression does not resemble any standard valuation form. This reflects its lack of financial priors and reliance on purely heuristic search. On the other hand, this might also demonstrate the importance of heuristic information in addition to reasoning capability in social science domains. 

\item In contrast, \textsc{mufasa}'s specialized agents recover adapted forms of earnings-multiple, discounted-cash-flow, book-value, relative-valuation, and residual-income models. This suggests that perspective disentanglement promotes economically grounded equation discovery, while allowing the learned forms to adapt to the data.

% Under non-stationary conditions, LLM-SR produced a linear combination of variables across different perspectives. This expression might be limited by its linear structure, compared to standard valuation identities, which typically rely on non-linear relationships. However, it could be the most effective strategy given that it was unable to split into multiple perspectives like ours.

% \item Our agents collectively recover adapted forms of various classical valuation models. The Earnings and Cash Flow agents resemble earnings and DCF-based valuations, the Asset agent aligns with book-value multiples, and the Quality agent reflects residual income intuition through ROE adjustments. This shows that disentangling the search into perspectives encourages the discovery of structurally meaningful and economically grounded equations. These equations were further adapted specifically to each dataset and aggregated by the meta-coordinator agent for each context. 
\end{itemize}

\paragraph{Released Artifacts.} 
The learnt equations for other datasets, the meta-coordinator weights, and the distilled agent learnings  $\mathbf{Z}^{(\mathbf{a})}$ are in Appendices \ref{app:expressions}, \ref{app:context_weights}, and \ref{app:agent_learnings}, respectively.

% $\mathbf{Z}^{(\mathbf{a})}$ can be found 
\section{Conclusion}
In this work, we study symbolic discovery for financial fundamental analysis, which constitutes a distinct class of symbolic regression problems. Unlike the natural sciences, financial valuation admits multiple valid perspectives, operates under non-stationary market conditions, and relies on noisy, continuous performance signals. To address this, we introduce \textsc{mufasa}, a hierarchical multi-agent framework that decomposes the discovery process into specialized agents, integrates them through hierarchical-level context reasoning, and leverages statistical-based memory to guide learning. We conduct extensive experiments across global markets and show that \textsc{mufasa} achieves strong forecasting performance while producing interpretable and economically grounded symbolic expressions. We release the learnt symbolic expressions and the distilled agent learnings to the public. 

% \paragraph{Limitations.}
% Some limitations of the proposed \textsc{mufasa} framework are discussed in Appendix \ref{app:limitations}.

\clearpage
\bibliography{aaai2027_long}

% \newpage

\newpage
\appendix
% Number appendix sections (A, A.1, ...) so \ref{...} in the main text
% resolves correctly; the main text itself stays unnumbered.
\setcounter{secnumdepth}{2}
\onecolumn
\AppendixTOC

\vspace{15px}
\section{Detailed Methods}
\label{detailed_methods}
Here, we provide the detailed algorithm for philosophy agent training and meta-coordinator training.

% \begin{wrapfigure}{r}{0.45\textwidth}
\begin{figure*}[!ht]
\vspace{-0.5\baselineskip}
\label{Algorithm_1}
\begin{minipage}{0.95\textwidth}
\footnotesize
\hrule
\vspace{0.4em}
\textbf{Algorithm 1: Philosophy Agent Training}
\vspace{0.4em}
\hrule
\vspace{0.4em}

\begin{algorithmic}[1]
\Require LLM $\pi_\theta$, database $\mathcal{X}t$, targets $y_{i,t+1}$, perspective $\mathbf{a}$, memory $\mathbf{M}^{(\mathbf{a})}$, learnings $\mathbf{Z}^{(\mathbf{a})}$

\State \textcolor{gray}{// Initialization}
\State $\tilde{f}^{(\mathbf{a})}, \mathbf{x}^{(\mathbf{a})}_{i,t} \gets \textsc{SeedEquation}(\mathbf{a})$
\State $\mathcal{L}^{(\mathbf{a})} \gets \mathrm{MAPE}\left(\tilde{f}^{(\mathbf{a})}(\mathbf{x}^{(\mathbf{a})}_{i,t}), y_{i,t+1}\right)$

\For{\textbf{all} iterations}
\State \textcolor{gray}{// Hypothesis Generation}
\State $\mathbf{p}^{(\mathbf{a})} \gets (\mathbf{a}, \tilde{f}^{(\mathbf{a})}, \mathbf{Z}^{(\mathbf{a})})$
\State $\tilde{f}^{(\mathbf{a})} \sim \pi\theta(\cdot \mid \mathbf{p}^{(\mathbf{a})})$

\State \textcolor{gray}{// Tool use: Data Retrieval}
\State $\mathbf{\hat{x}}^{(\mathbf{a})}_{i,t} \gets \textsc{Retrieve}\!\left( \mathcal{X}_t\right)$
\If{$\textsc{RetrieveError}$}
    \State $\mathbf{p}^{(\mathbf{a})} \gets \mathbf{p}^{(\mathbf{a})} \cup \textsc{ErrorFeedback}$
\EndIf

\State \textcolor{gray}{// Tool use: Code Evaluation}
\State $\hat{y}^{(\mathbf{a})}_{i,t+1} \gets \textsc{Evaluate}\!\left(\tilde{f}^{(\mathbf{a})}, \mathbf{\hat{x}}^{(\mathbf{a})}_{i,t}\right)$
\If{$\textsc{EvaluateError}$}
    \State $\mathbf{p}^{(\mathbf{a})} \gets \mathbf{p}^{(\mathbf{a})} \cup \textsc{ErrorFeedback}$
\EndIf

\State \textcolor{gray}{// Hypothesis Optimization}
\State $\hat{y}^{(\mathbf{a})}_{i,t+1} \gets \textsc{Optimize}\!\left(\hat{y}^{(\mathbf{a})}_{i,t+1}\right)$
\State $\mathcal{L}^{(\mathbf{a})} \gets \mathrm{MAPE}\!\left(\hat{y}^{(\mathbf{a})}_{i,t+1}, y_{i,t+1}\right)$
\State \textcolor{gray}{// Memory Update}
\State $\mathbf{M}^{(\mathbf{a})} \gets \mathbf{M}^{(\mathbf{a})} \cup \{(\tilde{f}^{(\mathbf{a})}, \mathcal{L}^{(\mathbf{a})})\}$

% \If{$\mathcal{L}^{(\mathbf{a})}_{k} < \mathcal{L}^{(\mathbf{a})}_{\star}$}
%     \State $\tilde{f}^{(\mathbf{a})}_{\star} \gets \tilde{f}^{(\mathbf{a})}_{k}$
%     \State $\mathcal{L}^{(\mathbf{a})}_{\star} \gets \mathcal{L}^{(\mathbf{a})}_{k}$
% \EndIf

\State $\mathbf{Z}^{(\mathbf{a})} \gets \textsc{UpdateLearnings}(\mathbf{M}^{(\mathbf{a})})$

\EndFor

\State \Return $\tilde{f}^{(\mathbf{a})}$
\end{algorithmic}
\vspace{0.2em}
\hrule
\end{minipage}
\vspace{-1.0\baselineskip}
% \vspace{-14px}
\end{figure*}
% \end{wrapfigure}

% \begin{wrapfigure}{r}{0.45\textwidth}
\begin{figure*}[!ht]
\vspace{-1.2\baselineskip}
% \vspace{-15px}
\label{Algorithm_2}
\begin{minipage}{0.95\textwidth}
\footnotesize
\hrule
\vspace{0.4em}
\textbf{Algorithm 2: Meta-Coordinator Training}
\vspace{0.4em}
\hrule
\vspace{0.4em}

\begin{algorithmic}[1]
\Require LLM $\pi_\phi$, specialist forecasts $\hat{y}^{(\mathbf{a})}_{i,t+1}$, targets $y_{i,t+1}$, memory $\mathbf{M}^{\text{meta}}$, learnings $\mathbf{Z}^{\text{meta}}$

\For{\textbf{all} $C(i,t)$}
    \State $\mathbf{\tilde{w}}_{C(i,t)} \gets \textsc{InitWeights}()$

    \For{\textbf{all} iterations}
        \State \textcolor{gray}{// Hypothesis Generation}
        \State $\mathbf{p}^{\text{meta}} \gets (C(i,t), \mathcal{L}^{(\mathbf{a})}, \mathbf{\tilde{w}}_{C(i,t)}, \mathbf{Z}^{\text{meta}})$
        \State $\mathbf{\tilde{w}}_{C(i,t)} \sim \pi_\phi(\cdot \mid \mathbf{p}^{\text{meta}})$

        \State \textcolor{gray}{// Tool use: Code Evaluation}
        \State $\hat{y}_{i,t+1} \gets \mathbf{\tilde{w}}_{C(i,t)}^\top \cdot\hat{\mathbf{y}}_{i,t+1}$
        \State $\mathcal{L}^{\text{meta}} \gets \mathrm{MAPE}\!\left(\hat{y}_{i,t+1}, y_{i,t+1}\right)$

        \State \textcolor{gray}{// Memory Update}
        \State $\mathbf{M}^{\text{meta}} \gets \mathbf{M}^{\text{meta}} \cup \{(\mathbf{\tilde{w}}_{C(i,t)}, \mathcal{L}^{\text{meta}}\}$
        \State $\mathbf{Z}^{\text{meta}} \gets \textsc{UpdateLearnings}(\mathbf{M}^{\text{meta}})$
    \EndFor
\EndFor

\State \Return $\mathbf{\tilde{w}}_{C(i,t)}$
\end{algorithmic}

\vspace{0.2em}
\hrule
\end{minipage}
\vspace{-1.0\baselineskip}
% \vspace{-5px}
\vspace{15px}
\end{figure*}
% \end{wrapfigure}

\newpage
\section{Statistical Significance Analysis}
\label{app:significance}
% We evaluate whether the performance improvements of \textsc{mufasa} over baselines are statistically significant using pairwise tests on prediction errors.

To assess the statistical significance of \textsc{mufasa}'s performance results, we apply the Diebold--Mariano (DM) test \citep{diebold2002comparing} under absolute percentage error loss. We incorporate two corrections.

First, we use a Newey--West heteroskedasticity and autocorrelation consistent (HAC) variance estimator \citep{newey1987simple} to account for serial correlation in the loss differential series across quarters. Specifically, we employ a Bartlett kernel with bandwidth selected according to Andrews' rule \cite{andrews1991heteroskedasticity}:
\begin{equation}
    L = \left\lfloor 4 \left(\frac{T}{100}\right)^{2/9} \right\rfloor,
\end{equation}
where $T$ denotes the number of test quarters. This adjustment is necessary because macroeconomic conditions that make one quarter difficult to forecast often persist, inducing autocorrelation in the loss differentials and leading to underestimated standard errors if uncorrected.

Second, we apply the Harvey--Leybourne--Newbold (HLN) small-sample correction \citep{harvey1997testing}, which adjusts the DM test statistic and evaluates it against a $t_{T-1}$ reference distribution rather than a standard normal distribution. This accounts for the small number of test quarters available per market.

% All tests are two-sided and conducted at the $\alpha = 0.05$ significance level.

\begin{table}[h]\footnotesize
\vspace{-10px}
\centering
\caption{Diebold--Mariano (DM) tests comparing \textsc{mufasa} against valuation and learning-based baselines. Entries report \textsc{mufasa} minus baseline in decimal form. Negative values indicate lower error (better performance). Significance: *** $p<0.001$, ** $p<0.01$, * $p<0.05$.}
\label{tab:dm_val_models}
\setlength{\tabcolsep}{4.0mm}
\begin{tabular}{lccccc}
\toprule
\textbf{Baseline} & \textbf{S\&P 500} & \textbf{Russell 2000} & \textbf{STOXX 600} & \textbf{FTSE Asia} & \textbf{CSI 300} \\
\midrule

\rowcolor{groupgray}\multicolumn{6}{c}{\textit{\textbf{Financial LLMs}}} \\
Fino1  
& -0.7283*** & -0.4515*** & -0.3067*** & -0.7925*** & -0.3408*** \\
Fin-R1 
& -0.1207*** & -0.0375*** & -0.1801*** & -0.5919*** & -0.4835*** \\

\rowcolor{groupgray}\multicolumn{6}{c}{\textit{\textbf{Financial Valuation}}} \\
P/E       
& -0.1628*** & -0.0371*** & -0.0691*** & -0.3519*** & -0.1158*** \\
EV/EBITDA 
& -0.1174*** & -0.1622*** & -0.0802*** & -0.2941*** & -0.0933*** \\
EV/EBIT   
& -0.1235*** & -0.0949*** & -0.0531*** & -0.2965*** & -0.0982*** \\
P/B       
& -0.2535*** & -0.0967*** & -0.1894*** & -0.1466*** & -0.0618*** \\
DDM       
& -0.2476*** & -0.0924*** & -0.1514*** & -0.1402*** & -0.0321** \\
FCFE      
& -0.2890*** & -0.2154*** & -0.1698*** & -0.1001** & -0.3787*** \\
FCFF      
& -0.2534*** & -0.1774*** & -0.1454*** & -0.2862*** & -0.2745*** \\

\rowcolor{groupgray}\multicolumn{6}{c}{\textit{\textbf{Symbolic Regression}}} \\
PySR      
& -0.2695*** & -0.0326*** & -0.0857*** & -0.3287*** & -0.0956*** \\
FunSearch 
& -0.3992*** & -0.3872*** & -0.1696*** & -0.4040*** & -0.1365*** \\
LLM-SR    
& -0.1101*** & -0.1751*** & -0.2345*** & -0.4354*** & -0.1442*** \\

\bottomrule
\end{tabular}
\vspace{-3px}
\end{table}

Across all pairwise comparisons, \textsc{mufasa} achieves statistically significant improvements in MAPE, indicating that the observed gains are mostly robust and not driven by random variations.
\newpage
\section{Learnt Symbolic Expressions}
\label{app:expressions}
We present the learnt symbolic expressions 
% across different datasets 
in Tables \ref{tab:learnt_equations_russell}--\ref{tab:learnt_equations_csi}.
Overall, we observe that \textsc{mufasa} consistently discovers structured and interpretable expressions that align with well-established financial valuation principles, while adapting them to data-specific and context-dependent variations.

A key observation is that the expressions discovered by \textsc{mufasa} closely resemble classical financial valuation formulas, such as discounted cash flow, book multiples and residual income models, as shown in the ``Theory'' column. 
Rather than directly reproducing these formulas, \textsc{mufasa} often recovers their core structural components (\eg scaling by cash flow or book value) and augments them with correction terms that reflect empirical patterns in the data.
These deviations from canonical formulas are expected in the financial setting. 
Unlike theoretical valuation models that rely on simplifying assumptions (\eg constant growth rates or discount factors), real-world financial data is often noisy, non-stationary, and subject to market-specific distortions. 
The additional multiplicative or additive adjustments learned by \textsc{mufasa} can thus be interpreted as data-driven refinements that account for such complexities, rather than arbitrary artifacts of overfitting.

In contrast, baseline methods such as PySR and FunSearch often produce expressions that are either overly complex or difficult to interpret, lacking clear correspondence to established financial theories. 
Similarly, LLM-SR capture only coarse additive relationships and fail to represent the non-linear structures commonly used in financial valuation. 
These comparisons highlight that \textsc{mufasa} strikes a balance between interpretability and flexibility, yielding expressions that are both theoretically grounded and empirically adaptive.
Importantly, we observe consistent structural patterns across different datasets (\eg Russell 2000, STOXX 600, FTSE Asia, and CSI 300), suggesting that the learnt symbolic forms capture generalizable valuation principles rather than dataset-specific artifacts.

% ===================== Russell 2000 =====================
\begin{table}[h]
\vspace{10px}
\centering
\resizebox{\textwidth}{!}{%
\begin{tabular}{llll}
\toprule
\textbf{Model} & \textbf{Equation} & \textbf{Theory Equation} & \textbf{Theory} \\
\midrule

Fino1 &
$\dfrac{EV/EBITDA\cdot(1+g_{EPS}/100)}{EV/EBITDA_{mkt}\cdot EBITDA}$ &
- &
- \\[0.4em]

Fin-R1 &
$EPS_{TTM}\cdot PE$ &
$EPS\cdot PE$ &
Earnings Multiple \\[0.4em]

PySR &
$(PB_{mkt,mean}+0.16)\cdot\left(\sqrt{|BVPS|+|e^{0.98}-BVPS|}+2.09-\dfrac{BVPS}{-0.55}-2.09\right)-0.29$ &
- &
- \\[0.4em]

FunSearch &
$2\cdot BVPS+10\cdot NI_{ps}$ &
- &
- \\[0.4em]

LLM-SR &
$0.11\cdot(PE-Y_{1})-0.01\cdot Y_{2}+0.20\cdot NI$ &
$R_i=\alpha+\beta_1X_{1,i}+\beta_2X_{2,i}+\cdots$ &
Linear Factor \\[0.4em]
\cdashline{1-4}
\rowcolor{groupgray}\multicolumn{4}{l}{\textsc{mufasa} agents} \\
Earnings &
$NI_{ps}\cdot PE\cdot\left(1+\dfrac{SustainGrowth}{100}\right)\cdot\left(1-1.12\cdot(0.19-1.35\times10^{-3}\cdot SustainGrowth)\right)$ &
$\dfrac{EPS}{r-g}$ &
Gordon Growth \\[0.4em]

Cashflow &
$FCF_{ps}\cdot EV/FCF\cdot\left(1+0.23\cdot(EV/EBITDA-1.00)\right)$ &
$\sum_{t=1}^{\infty}\dfrac{FCF_t}{(1+r)^t}$ &
Discounted Cash Flow \\[0.4em]

Asset &
$TBVPS\cdot PB\cdot\left(1-0.24\cdot(PB-1.00)\right)$ &
$BVPS\cdot PB$ &
Book Multiple \\[0.4em]

Growth &
$Revenue_{ps}\cdot\left(1+\dfrac{g_{Revenue}}{100}\right)$ &
$Revenue_0\cdot(1+g)$ &
Revenue Growth \\[0.4em]

Quality &
$EBITDA_{ps}\cdot P/FCF\cdot\left(1+\dfrac{EBITDA_{ps}-46.35}{100}\right)$ &
$BV+\sum_{t=1}^{\infty}\dfrac{(ROE_t-r)\cdot BV_{t-1}}{(1+r)^t}$ &
Residual Income \\

\bottomrule
\end{tabular}%
}
\caption{Learnt equations for the Russell 2000 dataset.}
\label{tab:learnt_equations_russell}
% \vspace{-10px}
\end{table}

% ===================== STOXX 600 =====================
\begin{table}[h]
\centering
\resizebox{\textwidth}{!}{%
\begin{tabular}{llll}
\toprule
\textbf{Model} & \textbf{Equation} & \textbf{Theory Equation} & \textbf{Theory} \\
\midrule

Fino1 &
$\dfrac{EV/EBITDA\cdot(1+g_{EPS}/100)}{EV/EBITDA_{mkt}\cdot EBITDA}$ &
- &
- \\[0.4em]

Fin-R1 &
$EPS_{TTM}\cdot PE$ &
$EPS\cdot PE$ &
Earnings Multiple \\[0.4em]

PySR &
$\left|
\dfrac{
\left((-0.99 + Div - (EBITDA + R\&D + |NI|) - NI)\cdot PB_{mkt,std}\right)
+ \left(EBITDA - (0.30 \cdot Rev + IntInc)\right)
}{Shares}
\right|
+ 0.68$ &
- &
- \\[0.4em]

FunSearch &
$0.25\cdot EBITDA_{ps}+0.35\cdot OCF_{ps}+0.20\cdot ROE+0.20\cdot BVPS_{lag4}$ &
- &
- \\[0.4em]

LLM-SR &
$-0.01\cdot PE+0.37\cdot PB+0.22\cdot(EV/EBIT)+0.36\cdot PB$ &
$R_i=\alpha+\beta_1X_{1,i}+\beta_2X_{2,i}+\cdots$ &
Linear Factor \\[0.4em]
\cdashline{1-4}
\rowcolor{groupgray}\multicolumn{4}{l}{\textsc{mufasa} agents} \\
Earnings &
$\dfrac{NI_{ps}\cdot PE\cdot(1+g_{EPS}/100)}{1+EBITDA_{ps}/100+(P/FCF/100)\cdot1.19}$ &
$\dfrac{EPS}{r-g}$ &
Gordon Growth \\[0.4em]

Cashflow &
$OCF_{ps}\cdot3.88\cdot\dfrac{P/FCF}{5.27}\cdot\left(1+0.02\cdot g_{EBITDA}\cdot\left(1-\dfrac{OCF_{ps}-6.66}{0.05}\right)\right)$ &
$\sum_{t=1}^{\infty}\dfrac{FCF_t}{(1+r)^t}$ &
Discounted Cash Flow \\[0.4em]

Asset &
$\dfrac{Assets-Liabilities}{Shares}\cdot\dfrac{PB}{28435.60+1.42\cdot LTInvest+5316.09\cdot(LTInvest/Assets-12.78)}$ &
$BVPS\cdot PB$ &
Book Multiple \\[0.4em]

Growth &
$Revenue_{ps}\cdot(1+g_{Revenue}/100)\cdot\left(1+\left(\dfrac{Revenue_{ps}}{809734.00}\right)^{34086.20}\right)\cdot\left(1-\left(1-\dfrac{EV/EBITDA}{100}\right)\cdot0.54\right)$ &
$Revenue_0\cdot(1+g)$ &
Revenue Growth \\[0.4em]

Quality &
$NI_{ps}\cdot(PE\cdot0.86)\cdot\left(1+\dfrac{OCF_{ps,lag1}/Assets_{lag4}}{100}\right)$ &
$BV+\sum_{t=1}^{\infty}\dfrac{(ROE_t-r)\cdot BV_{t-1}}{(1+r)^t}$ &
Residual Income \\

\bottomrule
\end{tabular}%
}
\caption{Learnt equations for the STOXX 600 dataset.}
\label{tab:learnt_equations_stoxx}
% \vspace{-10px}
\end{table}

% ===================== FTSE Asia =====================
\begin{table}[h]
\centering
\resizebox{\textwidth}{!}{%
\begin{tabular}{llll}
\toprule
\textbf{Model} & \textbf{Equation} & \textbf{Theory Equation} & \textbf{Theory} \\
\midrule

Fino1 &
$\dfrac{EV/EBITDA\cdot(1+g_{EPS}/100)}{EV/EBITDA_{mkt}\cdot EBITDA}$ &
- &
- \\[0.4em]

Fin-R1 &
$EPS_{TTM}\cdot PE$ &
$EPS\cdot PE$ &
Earnings Multiple \\[0.4em]

PySR &
$\dfrac{1}{\sqrt{e^{-1.27}}}\cdot
\log\left(
\left|
\dfrac{BVPS \cdot Debt_{ps}}
{-2.91\times10^{-3} - \dfrac{Debt_{ps} - InterestExpense}{R\&D + PB_{mkt,mean}}}
\cdot \dfrac{1}{\log(PB_{mkt,mean})}
\right|
+ 1.00
\right)$ &
- &
- \\[0.4em]

FunSearch &
$0.20\cdot EBITDA_{ps}+0.15\cdot BVPS_{lag4}+0.10\cdot ROE_{lag8}+0.10\cdot EBIT_{ps}+0.10\cdot EPS_{lag1}+0.05\cdot\dfrac{EV}{EBIT}$ &
- &
- \\[0.4em]

LLM-SR &
$5.86\times10^{-4}\cdot(EBIT-InterestExpense)-0.21\cdot R\&D-4.00\times10^{-3}\cdot GrossProfit+0.76\cdot BVPS$ &
$R_i=\alpha+\beta_1X_{1,i}+\beta_2X_{2,i}+\cdots$ &
Linear Factor \\[0.4em]
\cdashline{1-4}
\rowcolor{groupgray}\multicolumn{4}{l}{\textsc{mufasa} agents} \\
Earnings &
$NI_{ps}\cdot PE\cdot(1+0.001\cdot g_{EPS})\cdot\left(1-1.79\cdot\dfrac{\sigma_{EV/EBIT}}{100}\right)\cdot\left(1+0.03\cdot(\sigma_{EV/EBIT}>0.78)\right)$ &
$\dfrac{EPS}{r-g}$ &
Gordon Growth \\[0.4em]

Cashflow &
$OCF_{ps}\cdot\left(1+\dfrac{0.16\cdot g_{EBIT}}{100}\right)\cdot\left(\dfrac{EV/FCF}{(EV/FCF)_{mkt}}\cdot6.87\right)$ &
$\sum_{t=1}^{\infty}\dfrac{FCF_t}{(1+r)^t}$ &
Discounted Cash Flow \\[0.4em]

Asset &
$\dfrac{BVPS_{lag1}\cdot PB\cdot1.34}{PB}$ &
$BVPS\cdot PB$ &
Book Multiple \\[0.4em]

Growth &
$Revenue_{ps}\cdot(1+g_{Revenue}/100)\cdot\left(1+\dfrac{EV/EBITDA-(EV/EBITDA)_{mkt}}{(EV/EBITDA)_{mkt}}\right)$ &
$Revenue_0\cdot(1+g)$ &
Revenue Growth \\[0.4em]

Quality &
$NI_{ps}\cdot PE\cdot(1+ROE/100)$ &
$BV+\sum_{t=1}^{\infty}\dfrac{(ROE_t-r)\cdot BV_{t-1}}{(1+r)^t}$ &
Residual Income \\

\bottomrule
\end{tabular}%
}
\caption{Learnt equations for the FTSE Asia ex Japan dataset.}
\label{tab:learnt_equations_ftse}
% \vspace{-8px}
\end{table}

% ===================== CSI 300 =====================
\begin{table}[h!]
\centering
\resizebox{\textwidth}{!}{%
\begin{tabular}{llll}
\toprule
\textbf{Model} & \textbf{Equation} & \textbf{Theory Equation} & \textbf{Theory} \\
\midrule

Fino1 &
$\dfrac{EV/EBITDA\cdot(1+g_{EPS}/100)}{EV/EBITDA_{mkt}\cdot EBITDA}$ &
- &
- \\[0.4em]

Fin-R1 &
$EPS_{TTM}\cdot PE$ &
$EPS\cdot PE$ &
Earnings Multiple \\[0.4em]

PySR &
$\sqrt{\left|\dfrac{PB_{mkt,mean}\cdot0.10\cdot BVPS}{EBITDA-0.03+0.17\cdot EPS}+0.48+2.16\cdot Debt_{ps}\right|+\left|\dfrac{BVPS}{0.06-GrossProfit}\right|}+EPS$ &
- &
- \\[0.4em]

FunSearch &
$2\cdot BVPS+10\cdot NI_{ps}$ &
- &
- \\[0.4em]

LLM-SR &
$-0.02\cdot Debt_{ps}+1.53\times10^{-4}\cdot g_{Revenue}+0.54\cdot EV/Sales+0.52\cdot Revenue_{ps}$ &
$R_i=\alpha+\beta_1X_{1,i}+\beta_2X_{2,i}+\cdots$ &
Linear Factor \\[0.4em]
\cdashline{1-4}
\rowcolor{groupgray}\multicolumn{4}{l}{\textsc{mufasa} agents} \\
Earnings &
$NI_{ps}\cdot(1+g_{EPS}/100)\cdot(PE\cdot3.37)\cdot0.23$ &
$\dfrac{EPS}{r-g}$ &
Gordon Growth \\[0.4em]

Cashflow &
$0.09\cdot P/OCF+4.99\cdot OCF_{ps}\cdot(1+852.52\cdot(OCF_{ps}<0))\cdot(1+0.63\cdot(P/OCF>14.69))$ &
$\sum_{t=1}^{\infty}\dfrac{FCF_t}{(1+r)^t}$ &
Discounted Cash Flow \\[0.4em]

Asset &
$BVPS\cdot(PB-0.15\cdot PB_{mkt})\cdot1.00$ &
$BVPS\cdot PB$ &
Book Multiple \\[0.4em]

Growth &
$Revenue_{ps}\cdot(1+g_{Revenue}/100)$ &
$Revenue_0\cdot(1+g)$ &
Revenue Growth \\[0.4em]

Quality &
$EBIT_{ps}\cdot(1+ROE/100)\cdot P/FCF\cdot\left(1-0.94\cdot\left(\dfrac{Revenue_{ps}}{EBIT_{ps}}-1\right)\right)$ &
$BV+\sum_{t=1}^{\infty}\dfrac{(ROE_t-r)\cdot BV_{t-1}}{(1+r)^t}$ &
Residual Income \\

\bottomrule
\end{tabular}%
}
\caption{Learnt equations for the CSI 300 dataset.}
\label{tab:learnt_equations_csi}
% \vspace{15px}
\end{table}
\newpage
\section{Learnt Context Weighting}
\label{app:context_weights}
\begin{figure}[h]
\centering
\vspace{12px}
\makebox[\linewidth][c]{%
\resizebox{0.99\linewidth}{!}{%
\includegraphics[width=0.483\linewidth]{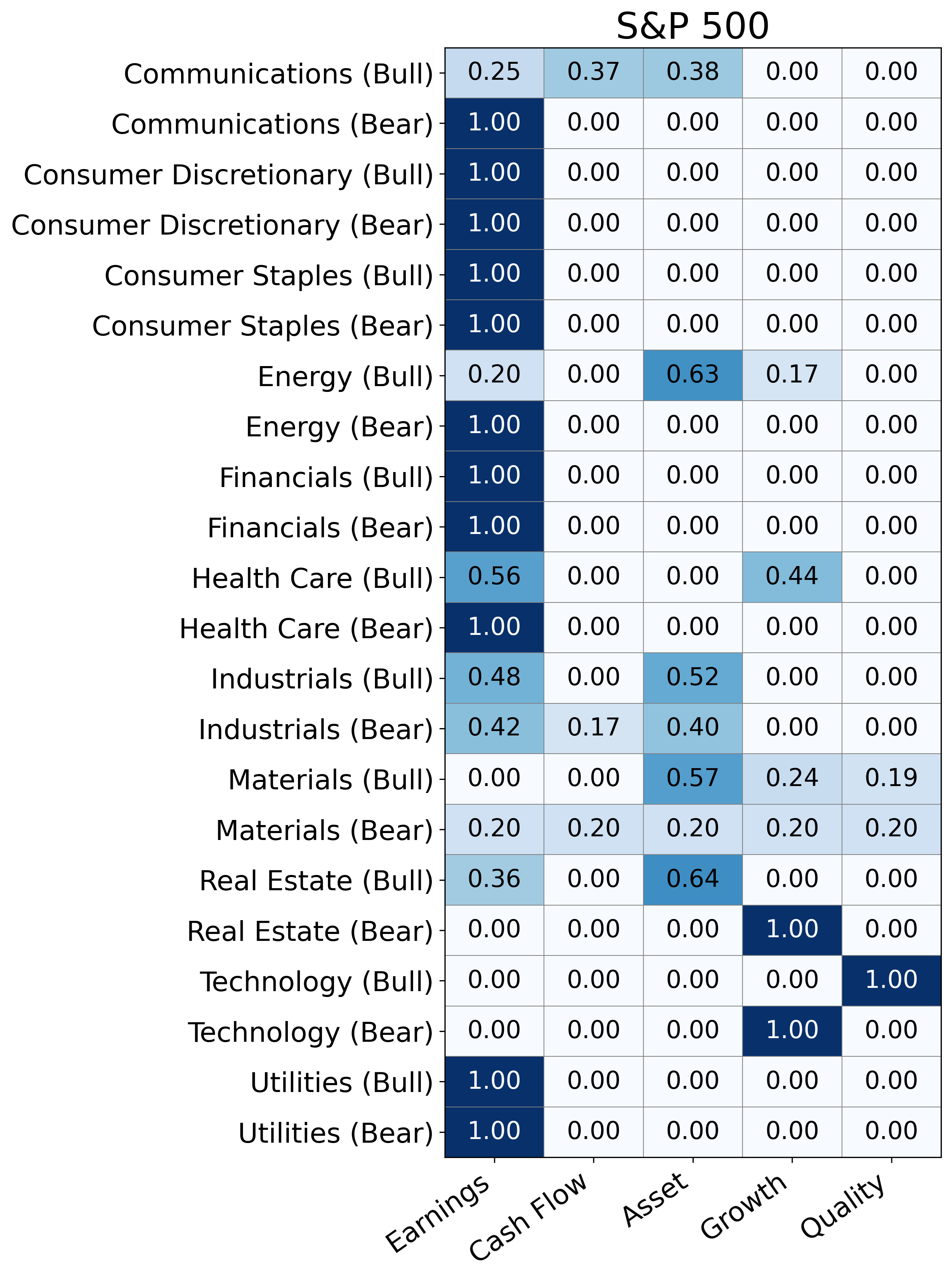}
\hspace{-5px}
\includegraphics[width=0.3\linewidth]{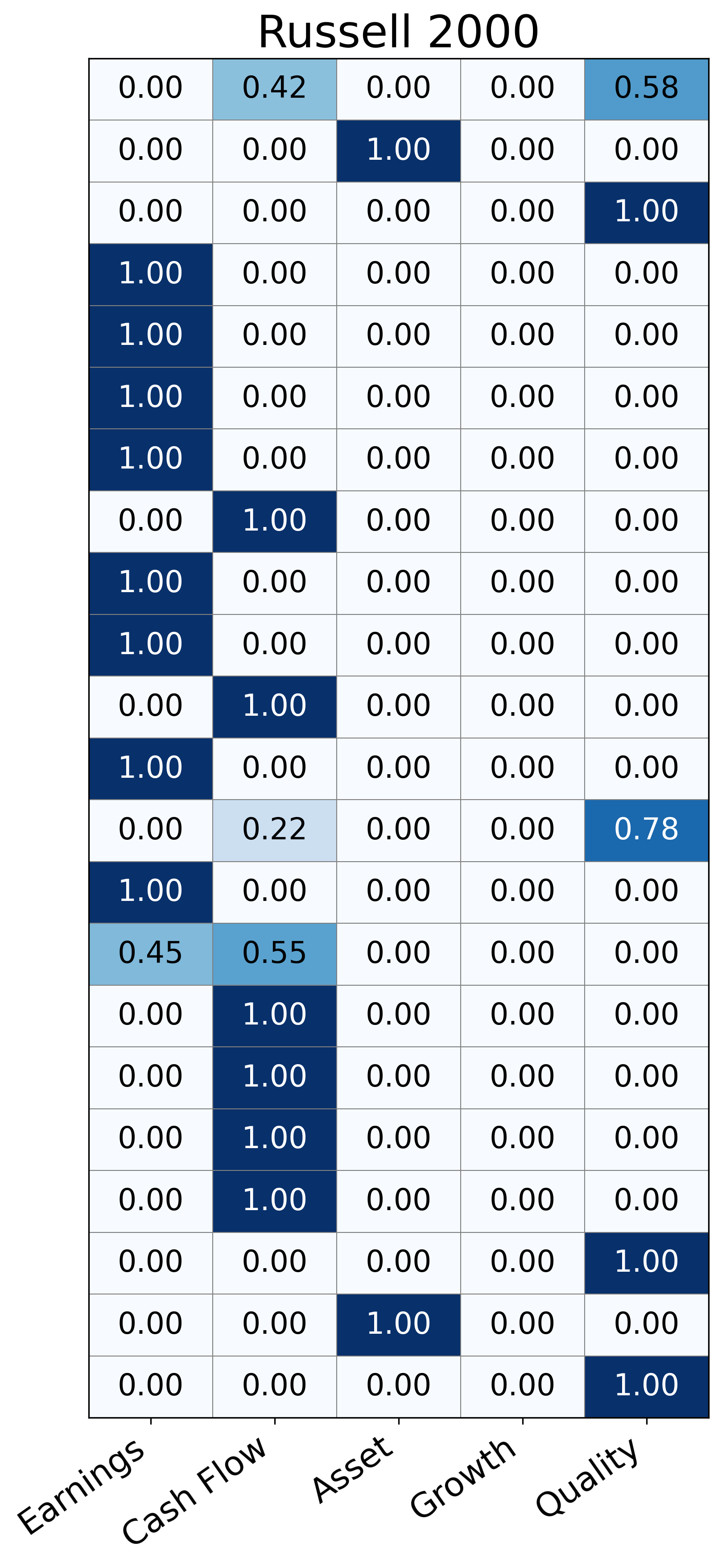}
\hspace{-5px}
\includegraphics[width=0.3\linewidth]{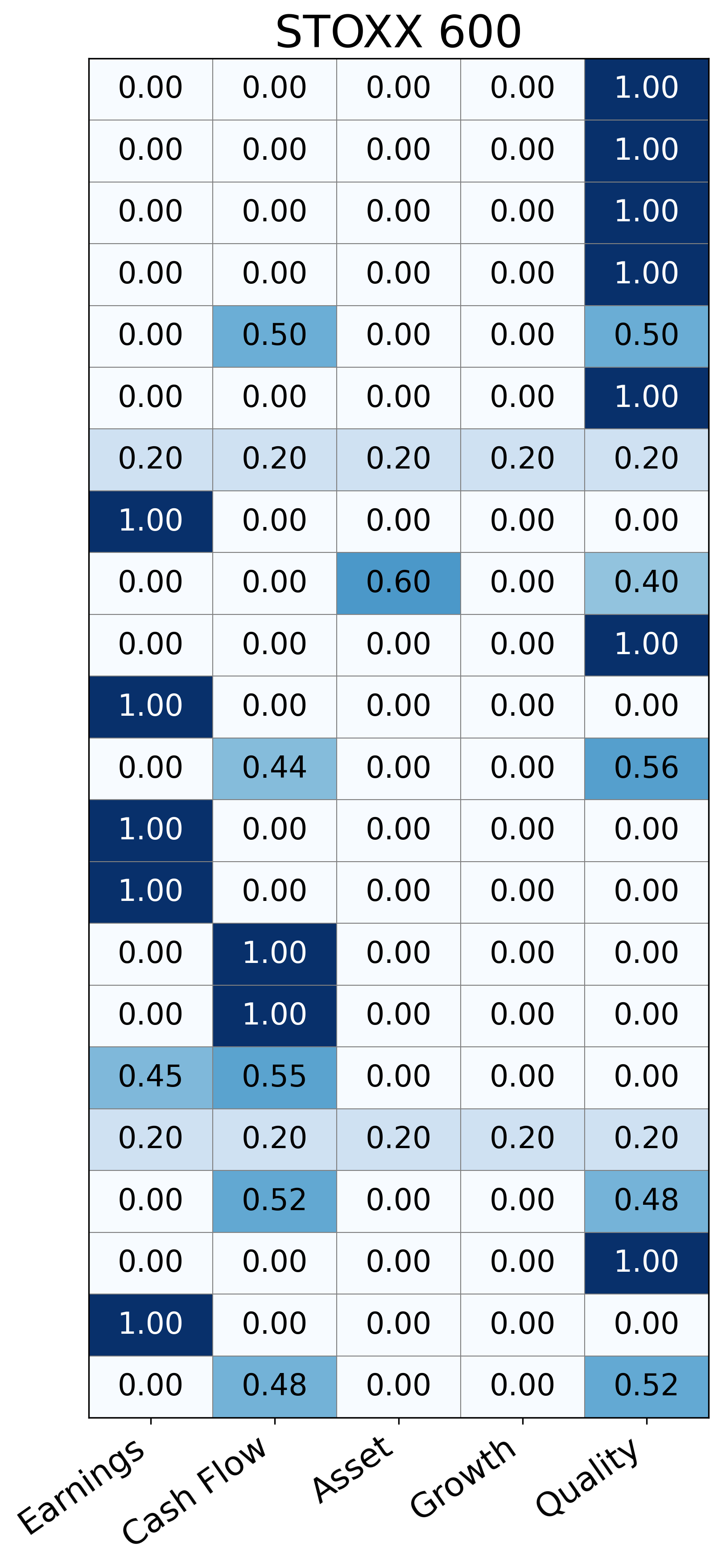}
}%
}

\vspace{0.3em}

\makebox[\linewidth][c]{%
\resizebox{0.66\linewidth}{!}{%
\includegraphics[width=0.483\linewidth]{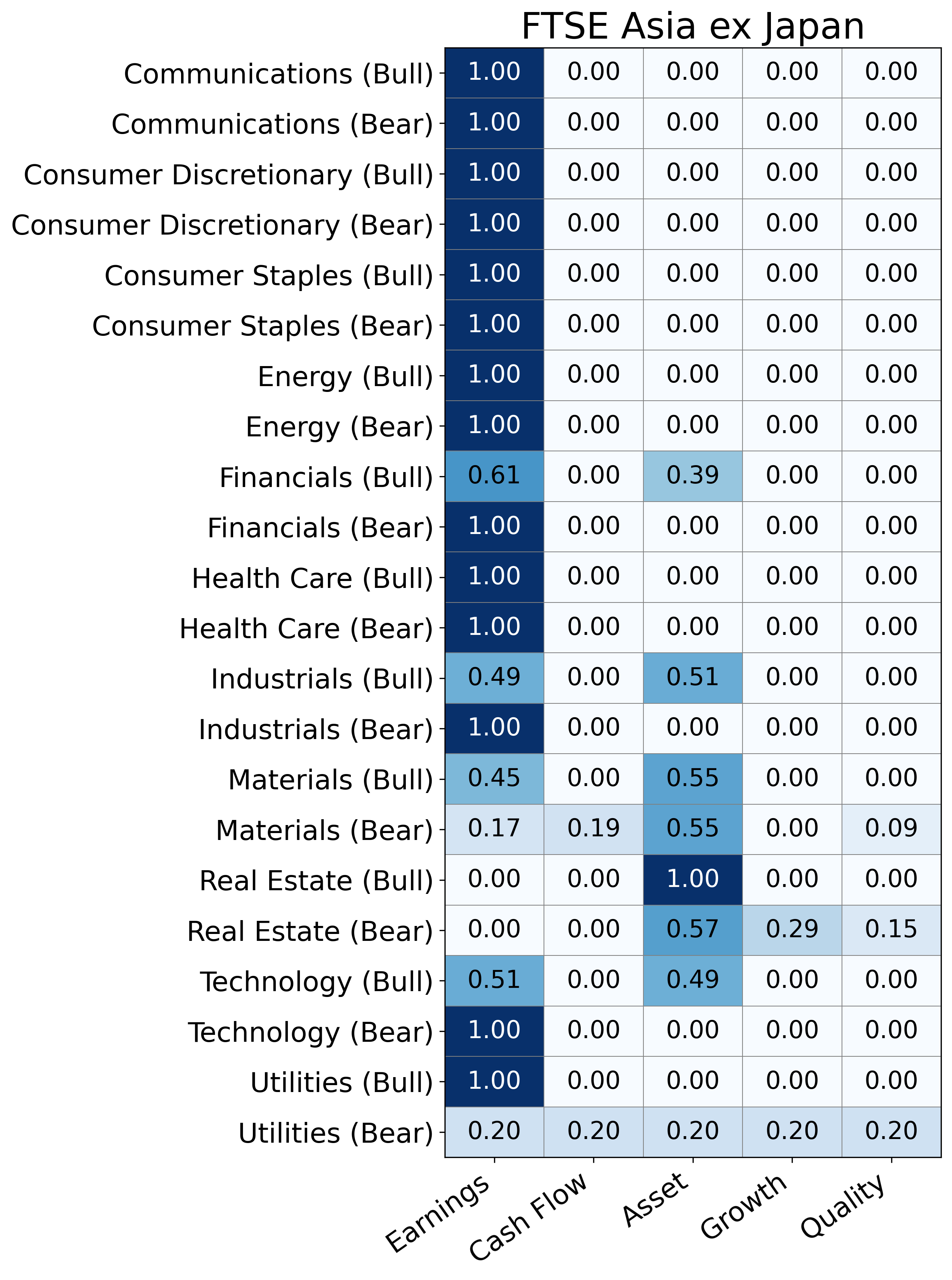}
\includegraphics[width=0.3\linewidth]{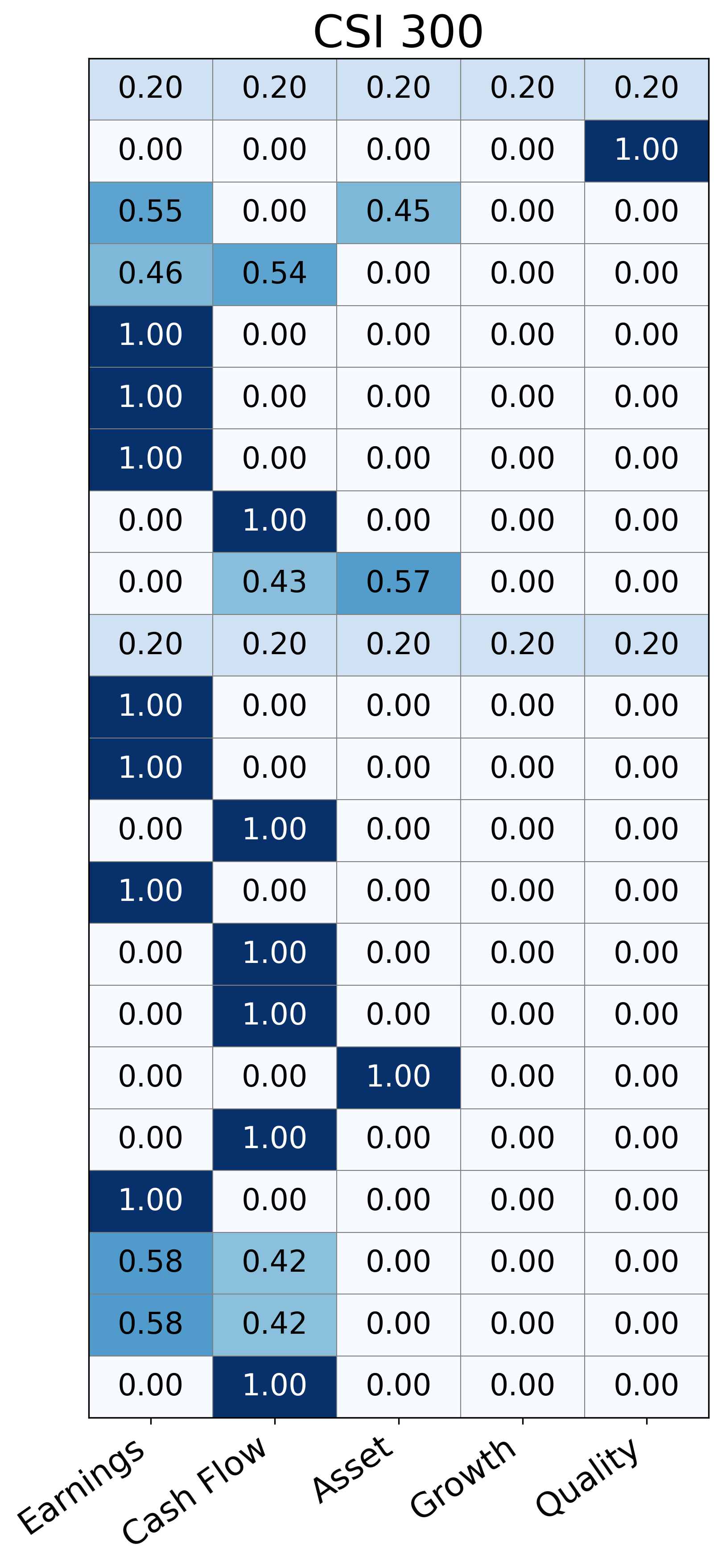}
}%
}

\caption{Learnt agent weighting across company sectors and market regimes for all datasets.}
\label{fig:weights_all}
\vspace{-30px}
\end{figure}

We illustrate the learnt context-dependent agent weighting over valuation perspectives in Figure~\ref{fig:weights_all} across the five datasets. 
Each heatmap shows how \textsc{mufasa} dynamically allocates importance to the different agent perspectives (\eg earnings, cash flow, asset, growth, and quality) 
conditioned on company sector and market regime (bull \textit{vs.}\ bear). 
Several consistent patterns emerge: 
\begin{itemize}[leftmargin=*]
\item First, the model assigns distinct weights across sectors, reflecting domain-specific valuation preferences. 
For example, asset- or cash flow-based signals are often emphasized in sectors with substantial balance sheet components (\eg real estate), while earnings-based signals are frequently assigned higher weights in sectors with more stable profitability profiles. 
% asset-based or cash flow-based perspectives tend to receive higher weights in asset-heavy sectors such as \textit{real estate} and \textit{utilities}, while earnings-based signals are more prominent in sectors with stable profitability. 
\item Second, the weights shift across bull and bear regimes, indicating that \textsc{mufasa} adapts its valuation strategy to changing market conditions. 
For example, it can be noticed that certain perspectives (\eg growth or earnings) are more emphasized in some bullish settings, while asset- or cash flow-based signals tend to receive relatively higher weights in some bearish contexts. 
These patterns are not uniform, but reflect a context-dependent reallocation of importance across valuation perspectives.

\item  Importantly, we observe recurring tendencies across geographically distinct markets, including the U.S. (S\&P 500, Russell 2000), Europe (STOXX 600), and Asia (FTSE Asia Pacific ex Japan, CSI 300), where certain valuation perspectives (\eg earnings or cash flow) are frequently assigned higher importance. 
While the exact weighting patterns vary across sectors and regimes, these recurring trends suggest that the learnt weighting mechanism is able to capture generalizable financial intuitions rather than purely dataset-specific artifacts.
\end{itemize}
Overall, the results demonstrate that the meta-coordinator effectively learns to combine multiple valuation perspectives across diverse market conditions, in a context-aware and interpretable manner.

\newpage
\clearpage
\section{Distilled Agent Learnings}
\label{app:agent_learnings}

%%% Earnings Agent %%%
\begin{tcolorbox}[
  enhanced, breakable,
  colback=white, colframe=black,
  boxrule=0.6pt, arc=1pt,
  left=4pt, right=4pt, top=4pt, bottom=4pt,
  title=\textbf{Earnings Agent Learnings},
  fonttitle=\bfseries,
]

\small

\textbf{Summary.}
Best structure uses earnings * PE with capped growth. The most promising direction is risk-adjusted sector growth caps. Avoid uncapped growth and raw PE multiples due to the persistent Technology sector gap.

\vspace{0.6em}
\textbf{Key Insights}
\begin{itemize}[leftmargin=1.2em, itemsep=0.15em]
\item Earnings-based structures require growth caps to avoid overprediction.
\item Sector-specific adjustments are necessary for high-PE sectors.
\item Raw PE-based formulations lead to systematic bias.
\item Growth signals improve performance only when bounded.
\end{itemize}

\vspace{0.6em}
\textbf{Statistical Diagnostics}
\begin{itemize}[leftmargin=1.2em, itemsep=0.15em]
\item Technology sector shows persistent overprediction relative to the global baseline.
\item High-PE sectors exhibit systematic bias in earnings-based models.
\item Growth variables without caps lead to unstable predictions.
\item Performance degrades when sector variation is not accounted for.
\end{itemize}

\vspace{0.6em}
\textbf{What Works}
\begin{itemize}[leftmargin=1.2em, itemsep=0.15em]
\item Using capped growth adjustments in earnings formulations.
\item Applying sector-specific growth caps.
\item Combining earnings with sector median PE.
\item Using bounded multiplicative structures, \eg growth caps.
\end{itemize}

\vspace{0.6em}
\textbf{What Fails}
\begin{itemize}[leftmargin=1.2em, itemsep=0.15em]
\item Using IS\_EPS\_GROWTH\_YOY without cap.
\item Using SECTOR\_MEDIAN\_PE without growth cap.
\item Using IS\_RD\_EXPEND\_TTM as a standalone multiplier.
\item Using uncapped growth-based formulations.
\end{itemize}

\vspace{0.6em}
\textbf{Important Variables}

\begin{center}
\setlength{\tabcolsep}{5pt}
\renewcommand{\arraystretch}{1.08}
\begin{tabular}{>{\raggedright\arraybackslash}p{0.42\linewidth} >{\raggedright\arraybackslash}p{0.50\linewidth}}
\toprule
\textbf{Variable} & \textbf{Observed Role} \\
\midrule
NET\_INCOME\_TTM\_PER\_SHARE & Core earnings driver. \\
SECTOR\_MEDIAN\_PE & Sector anchoring for valuation. \\
IS\_EPS\_GROWTH\_YOY & Growth signal requiring bounding. \\
IS\_RD\_EXPEND\_TTM & High-impact but unstable when standalone. \\
\bottomrule
\end{tabular}
\end{center}

\vspace{0.6em}
\textbf{Agent Adaptation Behavior}
\begin{itemize}[leftmargin=1.2em, itemsep=0.15em]
\item Introduces sector-specific growth caps after detecting overprediction.
\item Replaces uncapped growth terms with bounded structures.
\item Combines earnings with sector anchoring variables.
\item Moves away from standalone multipliers toward structured combinations.
\end{itemize}

\vspace{0.6em}
\textbf{Design Principle.}
Earnings-based models require sector-aware growth caps and bounded multiplicative structures to avoid systematic overprediction.

\end{tcolorbox}

\clearpage
%%% Cash Flow Agent %%%
\begin{tcolorbox}[
  enhanced, breakable,
  colback=white, colframe=black,
  boxrule=0.6pt, arc=1pt,
  left=4pt, right=4pt, top=4pt, bottom=4pt,
  title=\textbf{Cash Flow Agent Learnings},
  fonttitle=\bfseries,
]

\small

\textbf{Summary.}
Best structure uses cash flow with a sector-adjusted EV/EBITDA multiple. The most promising direction is a Consumer Discretionary discount factor. Avoid static multipliers without growth adjustments due to large sector gaps.

\vspace{0.6em}
\textbf{Key Insights}
\begin{itemize}[leftmargin=1.2em, itemsep=0.15em]
\item Sector-specific adjustments are required to correct extreme underprediction in Consumer Discretionary.
\item Growth-adjusted factors improve performance compared to static multipliers.
\item Cash flow quality adjustments are needed for early-stage companies with negative cash flow.
\item CF\_base * (1 + growth\_cap) structures outperform pure multiples by incorporating growth dynamics.
\end{itemize}

\vspace{0.6em}
\textbf{Statistical Diagnostics}
\begin{itemize}[leftmargin=1.2em, itemsep=0.15em]
\item Consumer Discretionary exhibits extreme underprediction driven by growth volatility.
\item Technology and Financials show systematic deviation due to differences in growth stability and regulatory complexity.
\item SECTOR\_MEDIAN\_EV\_EBITDA is unstable in asset-light companies due to denominator effects.
\item Ignoring negative cash flow via max() leads to systematic underestimation in early-stage firms.
\end{itemize}

\vspace{0.6em}
\textbf{What Works}
\begin{itemize}[leftmargin=1.2em, itemsep=0.15em]
\item Using sector-adjusted EV/EBITDA multiples.
\item Applying growth-adjusted scaling using SALES\_REV\_TURN\_TTM\_LAG4.
\item Incorporating cash flow quality adjustments for negative cash flow companies.
\item Using bounded structures, \eg CF\_base * (1 + growth\_cap).
\end{itemize}

\vspace{0.6em}
\textbf{What Fails}
\begin{itemize}[leftmargin=1.2em, itemsep=0.15em]
\item Using EBITDA\_YOY as a standalone multiplier.
\item Using static multipliers without growth adjustments.
\item Ignoring negative cash flow through truncation via max().
\end{itemize}

\vspace{0.6em}
\textbf{Important Variables}

\begin{center}
\setlength{\tabcolsep}{5pt}
\renewcommand{\arraystretch}{1.08}
\begin{tabular}{>{\raggedright\arraybackslash}p{0.42\linewidth} >{\raggedright\arraybackslash}p{0.50\linewidth}}
\toprule
\textbf{Variable} & \textbf{Observed Role} \\
\midrule
TRAIL\_12M\_FREE\_CASH\_FLOW\_PER\_SH & Core valuation base. \\
SECTOR\_MEDIAN\_EV\_EBITDA & Sector-adjusted valuation multiplier. \\
SALES\_REV\_TURN\_TTM\_LAG4 & Growth-adjustment factor with high sensitivity. \\
NEGATIVE\_CASH\_FLOW\_FACTOR & Adjustment for early-stage companies. \\
CONSUMER\_DISCRETIONARY\_FLAG & Sector-specific discount control. \\
\bottomrule
\end{tabular}
\end{center}

\vspace{0.6em}
\textbf{Agent Adaptation Behavior}
\begin{itemize}[leftmargin=1.2em, itemsep=0.15em]
\item Introduces sector-specific discount factors after detecting extreme underprediction.
\item Replaces static multipliers with growth-adjusted scaling.
\item Incorporates adjustments for negative cash flow rather than truncating values.
\item Proposes new combinations integrating growth, sector, and cash flow quality signals.
\end{itemize}

\vspace{0.6em}
\textbf{Design Principle.}
Cash flow models require sector-aware adjustments, growth scaling, and explicit handling of negative cash flow to avoid systematic underestimation.

\end{tcolorbox}

\clearpage
%%% Asset Agent %%%
\begin{tcolorbox}[
  enhanced, breakable,
  colback=white, colframe=black,
  boxrule=0.6pt, arc=1pt,
  left=4pt, right=4pt, top=4pt, bottom=4pt,
  title=\textbf{Asset Agent Learnings},
  fonttitle=\bfseries,
]

\small

\textbf{Summary.}
The best structure uses multiplicative adjustment for R\&D intensity and sector anchoring, avoiding invalid predictions and leveraging high-impact variables. The most promising direction is a Technology-specific R\&D multiplier. Avoid log-based adjustments and standalone BS\_CUR\_LIAB multipliers due to the persistent Technology sector gap.

\vspace{0.6em}
\textbf{Key Insights}
\begin{itemize}[leftmargin=1.2em, itemsep=0.15em]
\item Sector-specific multipliers are required to correct overprediction in the Technology sector.
\item R\&D intensity relative to assets is a high-impact adjustment factor.
\item Log-based adjustments introduce invalid predictions and should be avoided.
\item Liquidity adjustments improve stability in high-variance sectors.
\item max(BOOK\_VAL\_PER\_SH * x, BOOK\_VAL\_PER\_SH - y) * max(1, z) structures outperform pure multiples.
\end{itemize}

\vspace{0.6em}
\textbf{Statistical Diagnostics}
\begin{itemize}[leftmargin=1.2em, itemsep=0.15em]
\item Technology sector exhibits extreme overprediction due to overreliance on R\&D multipliers.
\item Log-based transformations result in invalid predictions.
\item BS\_CUR\_LIAB as a denominator causes invalid predictions in asset-light companies.
\item SECTOR\_MEDIAN\_P\_BOOK shows high volatility in Technology, leading to high CV and MAPE.
\item Consumer Staples shows underprediction due to underweighting of book value.
\end{itemize}

\vspace{0.6em}
\textbf{What Works}
\begin{itemize}[leftmargin=1.2em, itemsep=0.15em]
\item Using multiplicative adjustments for R\&D intensity.
\item Applying sector-specific adjustments using SECTOR\_MEDIAN\_P\_BOOK.
\item Using bounded structures, \eg max(1, ...) to avoid invalid predictions.
\item Incorporating liquidity buffer terms to reduce variance.
\end{itemize}

\vspace{0.6em}
\textbf{What Fails}
\begin{itemize}[leftmargin=1.2em, itemsep=0.15em]
\item Using BS\_CUR\_LIAB as a standalone multiplier.
\item Using log(1 + BOOK\_VAL\_PER\_SH / SECTOR\_MEDIAN\_P\_BOOK).
\item Using BS\_CUR\_LIAB as a denominator in asset-light companies.
\item Using unbounded R\&D multipliers in the Technology sector.
\end{itemize}

\vspace{0.6em}
\textbf{Important Variables}

\begin{center}
\setlength{\tabcolsep}{5pt}
\renewcommand{\arraystretch}{1.08}
\begin{tabular}{>{\raggedright\arraybackslash}p{0.42\linewidth} >{\raggedright\arraybackslash}p{0.50\linewidth}}
\toprule
\textbf{Variable} & \textbf{Observed Role} \\
\midrule
BOOK\_VAL\_PER\_SH & Core asset value and primary anchor. \\
SECTOR\_MEDIAN\_P\_BOOK & Sector anchoring variable. \\
BS\_CUR\_LIAB & Liquidity risk adjustment and variance driver. \\
IS\_RD\_EXPEND\_TTM & R\&D intensity has a high impact but is prone to overuse. \\
BS\_TOT\_ASSET & Scaling factor for asset normalization. \\
MARKET\_MEDIAN\_P\_BOOK & Market context variable with secondary importance. \\
\bottomrule
\end{tabular}
\end{center}

\vspace{0.6em}
\textbf{Agent Adaptation Behavior}
\begin{itemize}[leftmargin=1.2em, itemsep=0.15em]
\item Introduces Technology-specific multipliers after detecting overprediction.
\item Replaces log-based adjustments with bounded multiplicative structures.
\item Adds liquidity buffer terms to reduce high variance.
\item Explores combinations of R\&D intensity and sector anchoring variables.
\end{itemize}

\vspace{0.6em}
\textbf{Design Principle.}
Asset-based models require bounded multiplicative structures, sector-specific adjustments, and controlled use of high-impact variables to avoid instability and invalid predictions.

\end{tcolorbox}

\clearpage
%%% Growth Agent %%%
\begin{tcolorbox}[
  enhanced, breakable,
  colback=white, colframe=black,
  boxrule=0.6pt, arc=1pt,
  left=4pt, right=4pt, top=4pt, bottom=4pt,
  title=\textbf{Growth Agent Learnings},
  fonttitle=\bfseries,
]

\small

\textbf{Summary.}
The best structure uses MARKET\_MEDIAN\_P\_FCF with growth-capped FCF scaling. The most promising direction is a Consumer Discretionary discount factor. Avoid EBITDA\_YOY as a standalone multiplier due to persistent sector misfits.

\vspace{0.6em}
\textbf{Key Insights}
\begin{itemize}[leftmargin=1.2em, itemsep=0.15em]
\item Sector-specific discounts are required to correct extreme bias in Consumer Discretionary.
\item Growth elasticity improves performance compared to static multiples.
\item Piecewise growth structures improve flexibility over min/max formulations.
\item CF\_base * (1 + growth\_cap) structures improve MAPE by capturing growth dynamics.
\end{itemize}

\vspace{0.6em}
\textbf{Statistical Diagnostics}
\begin{itemize}[leftmargin=1.2em, itemsep=0.15em]
\item Consumer Discretionary exhibits extreme prediction error driven by FCF volatility.
\item Technology sector shows systematic overestimation due to growth effects.
\item TRAIL\_12M\_FREE\_CASH\_FLOW\_PER\_SH as a denominator causes invalid predictions.
\item SECTOR\_MEDIAN\_P\_FCF / MARKET\_MEDIAN\_P\_FCF shows high volatility in unstable sectors.
\end{itemize}

\vspace{0.6em}
\textbf{What Works}
\begin{itemize}[leftmargin=1.2em, itemsep=0.15em]
\item Using MARKET\_MEDIAN\_P\_FCF as a core multiplier.
\item Applying growth-capped scaling on free cash flow.
\item Using sector-specific scaling for Consumer Discretionary.
\item Replacing min/max structures with piecewise growth formulations.
\end{itemize}

\vspace{0.6em}
\textbf{What Fails}
\begin{itemize}[leftmargin=1.2em, itemsep=0.15em]
\item Using EBITDA\_YOY as a standalone multiplier.
\item Using SALES\_REV\_TURN\_YOY in multiplicative models.
\item Using TRAIL\_12M\_FREE\_CASH\_FLOW\_PER\_SH as a denominator.
\item Using SECTOR\_MEDIAN\_P\_FCF / MARKET\_MEDIAN\_P\_FCF in unstable sectors.
\end{itemize}

\vspace{0.6em}
\textbf{Important Variables}

\begin{center}
\setlength{\tabcolsep}{5pt}
\renewcommand{\arraystretch}{1.08}
\begin{tabular}{>{\raggedright\arraybackslash}p{0.42\linewidth} >{\raggedright\arraybackslash}p{0.50\linewidth}}
\toprule
\textbf{Variable} & \textbf{Observed Role} \\
\midrule
MARKET\_MEDIAN\_P\_FCF & Core multiplier for valuation. \\
TRAIL\_12M\_FREE\_CASH\_FLOW\_PER\_SH & Core growth-sensitive signal. \\
SECTOR\_MEDIAN\_P\_FCF & Sector adjustment factor. \\
MARKET\_MEDIAN\_P\_SALES & Alternative multiplier with instability in low-growth sectors. \\
EBITDA\_TTM\_PER\_SHARE & High-sensitivity variable with instability in high-growth sectors. \\
\bottomrule
\end{tabular}
\end{center}

\vspace{0.6em}
\textbf{Agent Adaptation Behavior}
\begin{itemize}[leftmargin=1.2em, itemsep=0.15em]
\item Introduces sector-specific discount factors after detecting extreme bias.
\item Replaces min/max structures with piecewise growth formulations.
\item Shifts from static multiples to growth-velocity models.
\item Explores combinations of growth scaling and sector adjustment.
\end{itemize}

\vspace{0.6em}
\textbf{Design Principle.}
Growth-based models require sector-specific adjustments, growth elasticity, and bounded scaling to avoid instability and extreme prediction errors.

\end{tcolorbox}

\clearpage
%%% Quality Agent %%%
\begin{tcolorbox}[
  enhanced, breakable,
  colback=white, colframe=black,
  boxrule=0.6pt, arc=1pt,
  left=4pt, right=4pt, top=4pt, bottom=4pt,
  title=\textbf{Quality Agent Learnings},
  fonttitle=\bfseries,
]

\small

\textbf{Summary.}
The best structure uses income * P\_FCF with growth adjustments. The most promising direction is a Technology-specific growth cap. Avoid EBITDA\_TTM\_PER\_SHARE as a standalone multiplier and fixed PE thresholds due to instability in growth regimes.

\vspace{0.6em}
\textbf{Key Insights}
\begin{itemize}[leftmargin=1.2em, itemsep=0.15em]
\item Sector-specific scaling is required to reduce bias in Technology.
\item Piecewise growth functions improve performance in high-return regimes.
\item Sector-agnostic growth caps improve stability and reduce overfitting.
\item CF\_base * (1 + growth\_cap) structures improve MAPE by balancing growth and stability.
\item Sector-agnostic growth modifiers improve performance when applied to income * P\_FCF structures.
\end{itemize}

\vspace{0.6em}
\textbf{Statistical Diagnostics}
\begin{itemize}[leftmargin=1.2em, itemsep=0.15em]
\item Technology sector shows bias due to the underestimation of growth.
\item Utilities show lower error driven by low growth expectations.
\item RETURN\_COM\_EQY causes invalid predictions in asset-light companies.
\item SECTOR\_MEDIAN\_P\_FCF introduces high CV in high-growth sectors.
\end{itemize}

\vspace{0.6em}
\textbf{What Works}
\begin{itemize}[leftmargin=1.2em, itemsep=0.15em]
\item Using income * P\_FCF as a base structure.
\item Applying piecewise growth functions instead of log growth.
\item Introducing sector-agnostic growth caps.
\item Using sector-specific scaling for high-growth sectors.
\end{itemize}

\vspace{0.6em}
\textbf{What Fails}
\begin{itemize}[leftmargin=1.2em, itemsep=0.15em]
\item Using EBITDA\_TTM\_PER\_SHARE as a standalone multiplier.
\item Using SECTOR\_MEDIAN\_PE with fixed thresholds.
\item Using RETURN\_COM\_EQY in unstable regimes without constraints.
\end{itemize}

\vspace{0.6em}
\textbf{Important Variables}

\begin{center}
\setlength{\tabcolsep}{5pt}
\renewcommand{\arraystretch}{1.08}
\begin{tabular}{>{\raggedright\arraybackslash}p{0.42\linewidth} >{\raggedright\arraybackslash}p{0.50\linewidth}}
\toprule
\textbf{Variable} & \textbf{Observed Role} \\
\midrule
OPER\_CF\_TTM\_PER\_SHARE & Strong signal when combined with P\_FCF. \\
SECTOR\_MEDIAN\_P\_FCF & Core multiplier with sector adjustment role. \\
RETURN\_COM\_EQY & Growth signal with instability in high-return regimes. \\
NET\_INCOME\_TTM\_PER\_SHARE & Core valuation base. \\
SECTOR\_MEDIAN\_PE & Alternative multiplier with overfitting risk. \\
EBITDA\_TTM\_PER\_SHARE & High-volatility variable with unstable performance. \\
\bottomrule
\end{tabular}
\end{center}

\vspace{0.6em}
\textbf{Agent Adaptation Behavior}
\begin{itemize}[leftmargin=1.2em, itemsep=0.15em]
\item Introduces sector-specific scaling after detecting bias in Technology.
\item Replaces log growth with piecewise linear growth functions.
\item Adds sector-agnostic growth caps to reduce overfitting.
\item Explores combinations of income, growth, and sector multipliers.
\end{itemize}

\vspace{0.6em}
\textbf{Design Principle.}
Quality-based models require growth-aware scaling, bounded transformations, and sector-specific adjustments to balance stability and predictive performance.

\end{tcolorbox}

\newpage
\paragraph{Analysis of Agent Learnings.}
We highlight consistent patterns observed across the learnings:

\begin{itemize}[leftmargin=*]
    \vspace{-5px}
    \item Many candidate formulations fail not due to poor average accuracy, but due to instability or invalid behavior under specific conditions. This is reflected in the statistical diagnostics such as high variance, high CV, or invalid outputs arising from denominator effects (\eg cash flow or liability terms in asset-light firms). Such failure modes would not be identifiable from scalar loss alone, highlighting the necessity of the statistical memory in guiding the search toward robust expressions.

    \item Instability is consistently conditional on firm type and sector interactions rather than global. For instance, sector-based multiples exhibit high volatility in certain sectors, and variables such as free cash flow or liabilities lead to breakdowns in early-stage or asset-light companies. This explains why a single global expression is insufficient, and also motivates the use of context-dependent aggregation across specialized agents.
    
    \item A recurring observation is the presence of systematic over- and underestimation across sectors. For example, Technology frequently exhibits persistent overprediction (earnings, asset, growth agents), while Consumer Discretionary shows extreme underprediction or volatility-driven errors (cash flow, growth agents). These directional biases justify the need for sector- and regime-aware weighting.
        
    \item Variables are rarely discarded outright, but instead require structured use to remain effective. Several high-impact features (\eg growth signals, R\&D intensity, sector multiples) are repeatedly identified as unstable or biased when used in isolation, but become useful when combined with anchoring or adjustment terms. This indicates that the statistical diagnostics not only identify failure modes, but also guide the construction of stable compositions within each valuation perspective.
\vspace{10px}
\end{itemize}
% \input{e_agent_prompts}
% \clearpage
\vspace{-8px}
\section{Extended Ablation Study}
\label{app:extended_ablation}
\begin{table*}[h]
\footnotesize
\centering
\begin{tabular}{lccccc}
\toprule
\textbf{Model Variant} & \textbf{S\&P 500} & \textbf{Russell 2000} & \textbf{STOXX 600} & \textbf{FTSE APAC} & \textbf{CSI 300} \\
\midrule

\multicolumn{6}{c}{\textbf{Binary Memory (Pass/Fail Signals Only)}} \\
\midrule
Single-Perspective Agent 
& 0.5699 & 0.7527 & 0.7220 & 0.6095 & 0.5735 \\

Multi-Agent (Equal Weights) 
& 0.5466 & 1.0459 & 0.6454 & 0.7088 & 0.5599 \\

Multi-Agent (Best Single Agent) 
& 0.5127 & 1.5160 & 0.6212 & \textbf{0.6083} & 0.5761 \\

Multi-Agent (Meta-Coordinator) 
& \textbf{0.4646} & \textbf{0.6445} & \textbf{0.5759} & 0.6086 & \textbf{0.5070} \\

\midrule
\multicolumn{6}{c}{\textbf{MAPE-Only Memory (No Additional Statistics)}} \\
\midrule
Single-Perspective Agent 
& 0.4983 & 0.7208 & 0.5858 & 0.5947 & 0.5517 \\

Multi-Agent (Equal Weights) 
& 0.5333 & 0.9027 & 0.5881 & 0.6393 & 0.5175 \\

Multi-Agent (Best Single Agent) 
& 0.5127 & \textbf{0.6948} & 0.6212 & 0.5926 & 0.5929 \\

Multi-Agent (Meta-Coordinator) 
& \textbf{0.4702} & 0.7525 & \textbf{0.5398} & \textbf{0.5852} & \textbf{0.4852} \\

\midrule
\multicolumn{6}{c}{\textbf{Full Statistical Memory (Variance, Tail Risk, \etc)}} \\
\midrule
Single-Perspective Agent 
& 0.4971 & 0.8987 & 0.7608 & 0.6095 & 0.5928 \\

Multi-Agent (Equal Weights) 
& 0.4652 & 0.8831 & 0.5448 & 0.6442 & 0.5309 \\

Multi-Agent (Best Single Agent) 
& 0.4676 & 1.9170 & 0.5256 & 0.5833 & 0.4739 \\

\textsc{mufasa} (Ours) 
& \textbf{0.4465} & \textbf{0.5718} & \textbf{0.5064} & \textbf{0.5814} & \textbf{0.4343} \\

\bottomrule
\end{tabular}
\caption{Extended ablation study. We compare strategies under different memory settings.}
\label{tab:ablation_full}
\end{table*}

We extend the ablation study on the statistical agent memory in Table \ref{tab:ablation_full}. \textsl{Single-Perspective Agent} uses only a single LLM agent to search for the best expression, \textsl{Multi-Agent (Equal Weights)} use multiple agents without hierarchical-level coordination, \textsl{Multi-Agent (Best Single Agent)} reports the best post-hoc performance out of the multiple agents. Across all memory settings, \textsl{Multi-Agent (Meta-Coordinator)} generally achieves the strongest performance, often outperforming all of the best individual agents. This suggests that the gain does not come simply from finding the single strongest perspective in the agent pool, but from learning how to combine the complementary perspectives.
% The best single agent can perform well in certain markets, but its performance is less stable and depends on knowing which perspective is most suitable after evaluation. In contrast, the meta-coordinator provides a more practical and robust mechanism by adaptively weighting agents based on past experience.

Across the memory variants, \textsl{Binary Memory} refers to storing past equations with binary pass/fail signals only, \textsl{MAPE-Only Memory} stores their MAPE performance, while \textsl{Full Statistical Memory} stores the full suite of statistical information listed in Table \ref{tab:memory_stats}. We find that using the full statistical memory typically result in the strongest performance across all model variants. Full statistical memory provides information about not only the average error, but also the reliability and risk profile of each agent's expressions. This allows the agents to distinguish equations that perform well consistently and those that perform well only occasionally, resulting in more robust expressions.

Overall, the results show that \textsc{mufasa} benefits from using multi-perspective agents, hierarchical-level coordination and richer statistical feedback, leading to more consistent symbolic discovery across markets. We do a deeper analysis of the statistical memory performance in the following section.
\section{Additional Statistical Results}
\label{app:additional_results}
In Table \ref{tab:memory_stats}, we show that the full statistical memory stores information such as Spearman Rank Correlation, Percentile Errors, and Mean Signed Percentage Error (MSPE), among others. We perform a deeper analysis of the different statistical performances of the learnt symbolic expressions. 

\begin{figure}[h]
% \vspace{-5px}
\centering
\begin{subfigure}{0.24\columnwidth}
    \includegraphics[width=\linewidth]{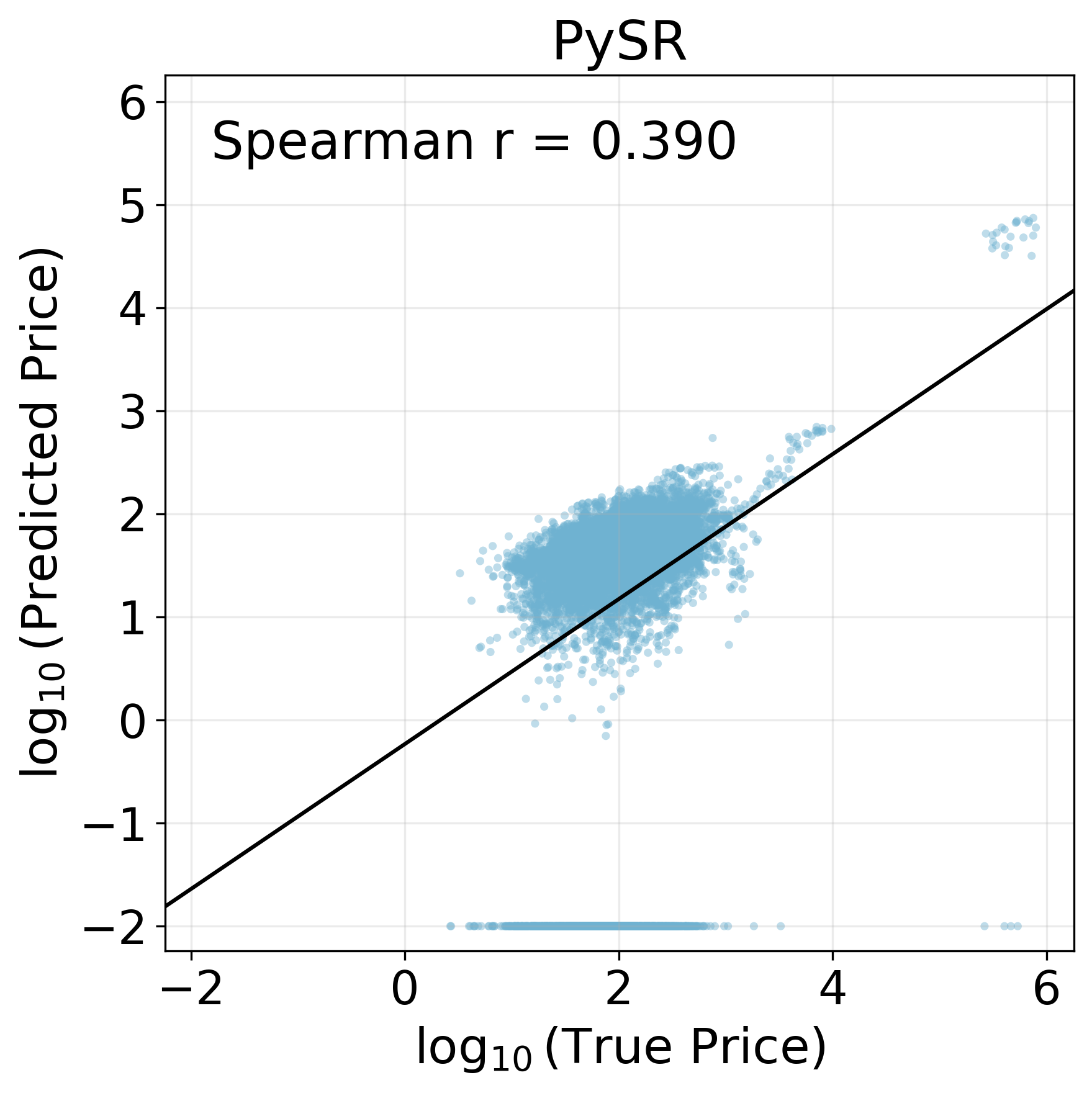}
\end{subfigure}
\begin{subfigure}{0.24\columnwidth}
    \includegraphics[width=\linewidth]{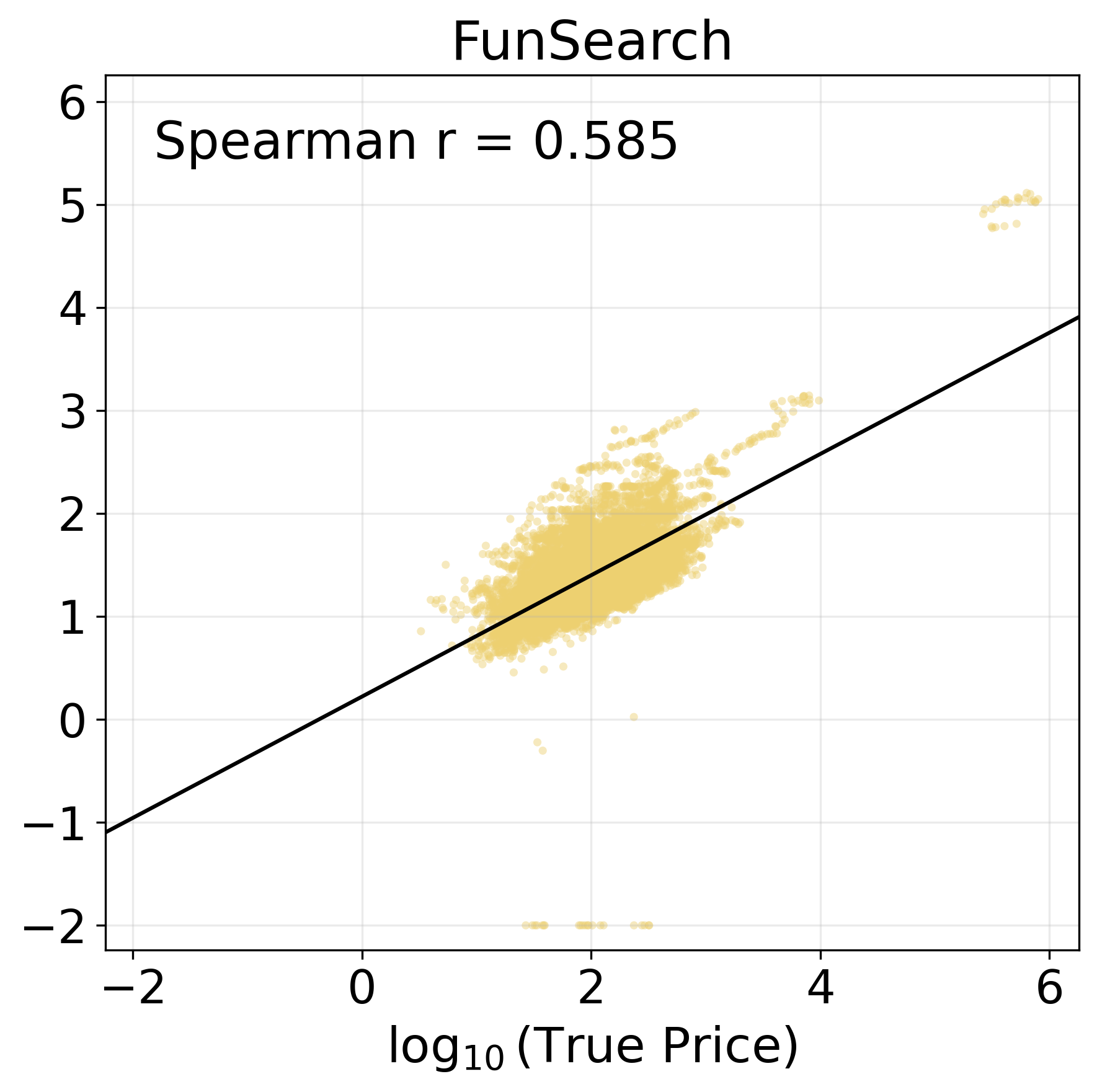}
\end{subfigure}
\begin{subfigure}{0.24\columnwidth}
    \includegraphics[width=\linewidth]{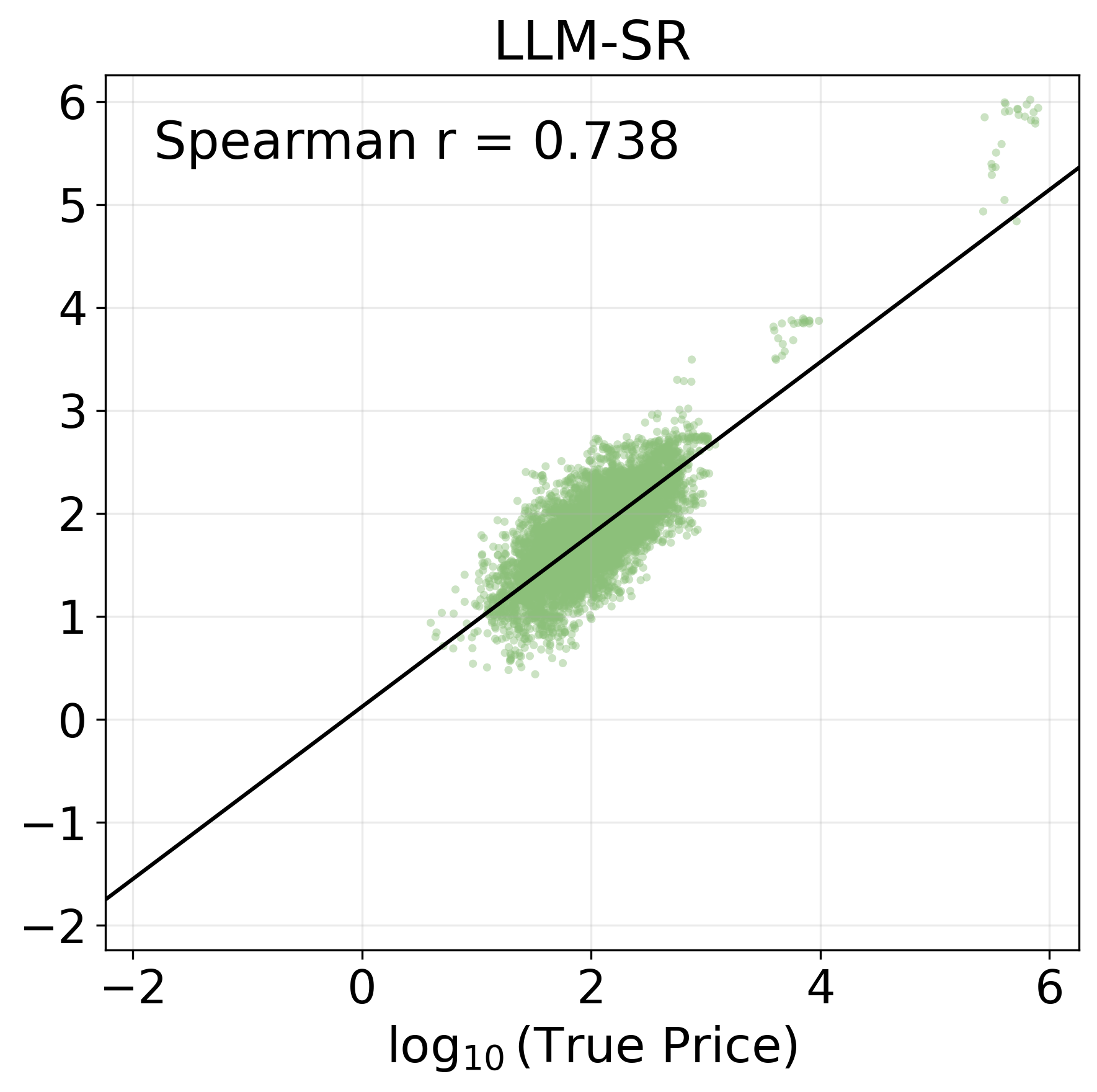}
\end{subfigure}
\begin{subfigure}{0.24\columnwidth}
    \includegraphics[width=\linewidth]{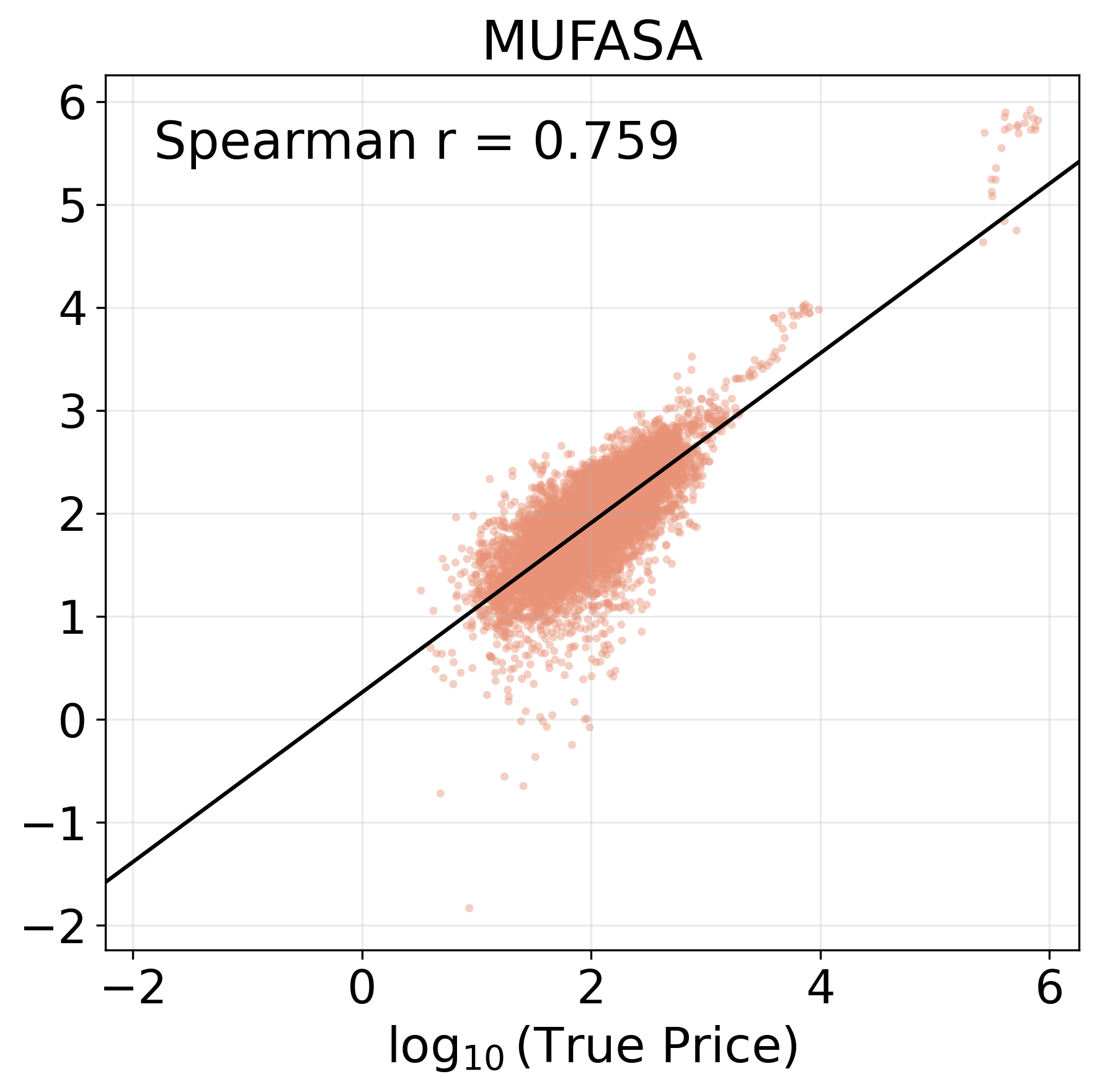}
\end{subfigure}
\caption{Predicted \textit{vs.} True log prices across Symbolic Regression models on the S\&P 500 dataset.}
\vspace{-10px}
\label{fig:spearman}
\end{figure}

\paragraph{Spearman Correlation.} 
Many of the learnt symbolic expressions are non-linear and multiplicative (see Table \ref{tab:learnt_equations}), and the MAPE metric might not be fully indicative of their quality. For example, a mismatched multiplier would result in high MAPE, but could still be correlated with the true symbolic relationship. We additionally analyze the Spearman correlation between the predicted and true prices.

Notably, in Figure \ref{fig:spearman}, we find that MAPE and rank correlation capture different aspects of performance: PySR achieved better MAPE than FunSearch in the S\&P 500 dataset (see Table \ref{tab:results}) but shows lower rank correlation. This might suggest that its heuristic search methodology fits average values without preserving meaningful structural relationships, in contrast to FunSearch, which uses LLM reasoning.

We see fewer outlier points going from left to right, given that PySR simply fits average MAPE, while LLM methods like FunSearch search for structural expressions but only evaluate them over the MAPE metric. Our memory structure captures a more holistic set of statistical performances, which includes the rank correlation metric. Overall, \textsc{mufasa} achieved the highest Spearman correlation.  

\begin{figure}[h]
\includegraphics[width=\columnwidth]{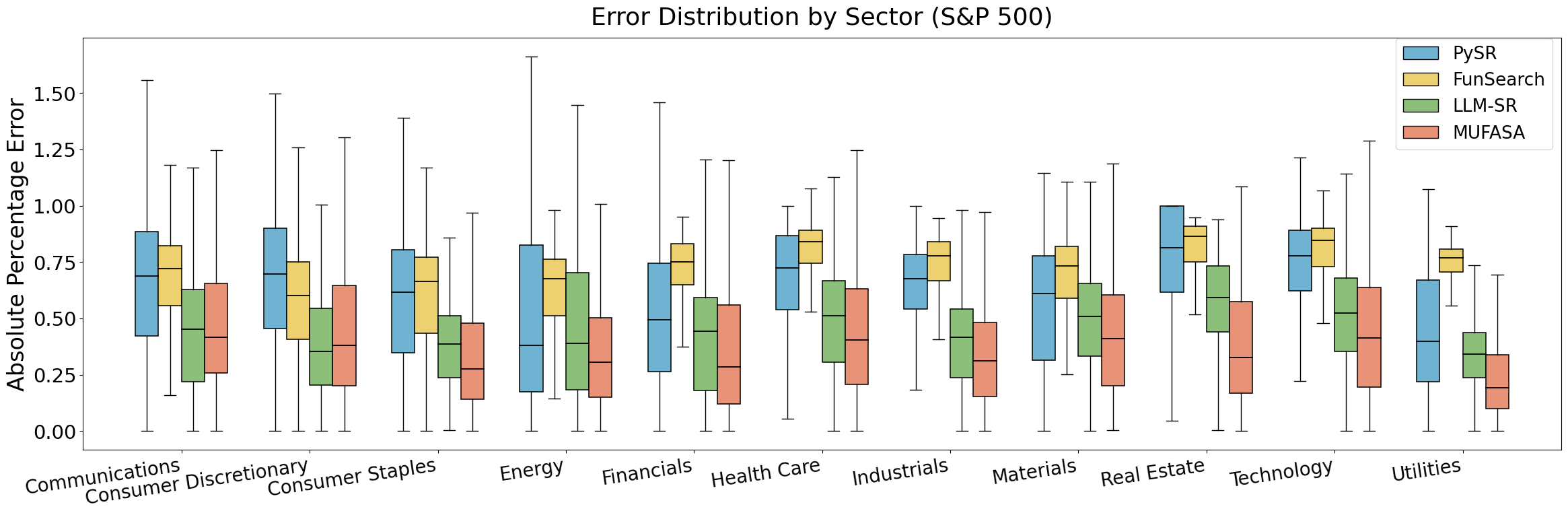}
\caption{Error distribution across sectors. The center lines show median errors, while the top and bottom edges of the box plots show the 25\textsuperscript{th} and 75\textsuperscript{th} percentile errors, respectively. Lower is better.}
\label{fig:mufasa-boxplot}
\end{figure}

\paragraph{Error Percentiles.} In Figure \ref{fig:mufasa-boxplot}, we do a deeper analysis using a breakdown by company sectors. 

\textsc{mufasa} achieves the strongest performance in sectors such as Utilities, Industrials, Real Estate, and Consumer Staples, where error distributions are both lower and more concentrated. These sectors are typically characterized by their fundamentals, making them well-aligned with the accounting information used by the model. The stability of these sectors allows the model to effectively perform symbolic discovery without being disrupted by the high variability of the stock prices used in evaluation.

% \begin{wraptable}{r}{0.50\textwidth}
\begin{figure}[!ht]
\vspace{-8px}
\footnotesize
\centering
\vspace{-0.5em}
\resizebox{0.50\textwidth}{!}{
\begin{tabular}{lccc}
\toprule
Model & 25\textsuperscript{th} pct. ($\downarrow$) & Median ($\downarrow$) & 75\textsuperscript{th} pct. ($\downarrow$) \\
\midrule
PySR      & 0.4103 & 0.6249 & 0.8322 \\
FunSearch & 0.6136 & 0.7500 & 0.8283 \\
LLM-SR    & 0.2663 & 0.4475 & 0.6086 \\
\textsc{mufasa} & \textbf{0.1725} & \textbf{0.3382} & \textbf{0.5556} \\
\bottomrule
\end{tabular}
}
\caption{Aggregate APE distributions.}
\label{tab:ape_distribution_summary}
\vspace{-0.8em}
\end{figure}
% \end{wraptable}

While \textsc{mufasa} improves performance across most sectors, it shows slightly higher upper- and lower-percentile errors in the Communications and Consumer Discretionary sectors. A possible explanation is that these sectors are less directly driven by fundamental signals and are more influenced by forward-looking expectations, intangible assets, and market sentiment. As a result, it could fail to capture unexpected jumps caused by news events, which were not considered in our work. However, our median error is still lower or close to benchmark performance in these sectors.

In general, \textsc{mufasa} achieves the lowest averaged errors across all percentiles of the distribution, including the median, as shown in Table \ref{tab:ape_distribution_summary}. Similarly, our memory structure (Table \ref{tab:memory_stats}) also captures and learns from tail errors, allowing it to produce more robust expressions compared to baselines.

\paragraph{Directional Bias.}
\begin{figure}[h]
\includegraphics[width=\columnwidth]{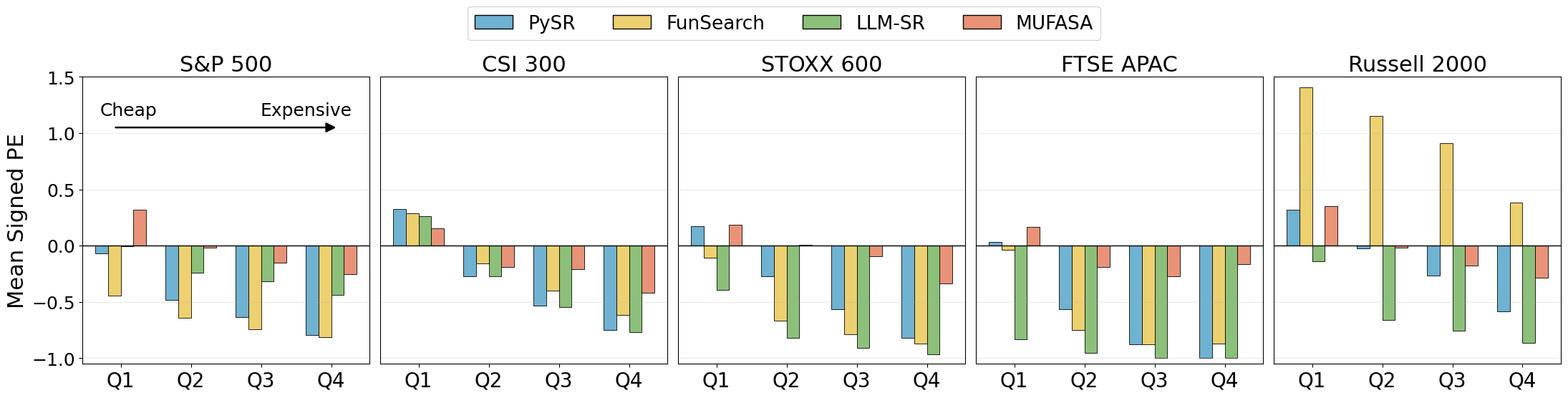}
\caption{Mean Signed Percentage Error across valuation quartiles (Q1: Cheap to Q4: Expensive).}
\label{fig:mspe}
\end{figure}
In Figure \ref{fig:mspe}, we further analyze directional bias across valuation quartiles, using the mean signed percentage error (MSPE) metric. Across all markets, we find that \textsc{mufasa} tends to overestimate value for cheaper companies (Q1) and underestimate value for expensive ones (Q4).

However, compared to baselines, \textsc{mufasa} exhibits lower directional bias overall, with errors remaining closer to zero across all quartiles. This indicates that the model not only improves absolute accuracy (as seen in MAPE), but also avoids systematic mispricing across different valuation regimes. 

Notably, the bias patterns are also more stable across markets for \textsc{mufasa}, while competing methods display greater variability. This might be attributed to the statistical memory, which also tracks bias-related signals such as MSPE, enabling the agent to reduce systematic over- or under-prediction across the different price regimes. These could be seen in the distilled agent learnings (see Statistical Diagnostics in Section \ref{app:agent_learnings}), which often mention observations on over- and under-estimations.

% Overall, these results reinforce that \textsc{mufasa} improves both predictive accuracy and directional consistency, aligning with our objective of learning robust symbolic expressions under noisy and heterogeneous financial data.
\vspace{15px}
\section{Limitations of the work}
\label{app:limitations}

The evaluation of classical valuation baselines depends on a set of modeling assumptions that are standard in finance, but not uniquely defined. For example, our baseline methods require specifying how future cash flows or dividends evolve over time \cite{damodaran2012investment}. In this work, we adopt commonly used approaches such as the constant-growth assumption \cite{gordon1962savings} for consistency and interpretability. Importantly, evaluating across all possible assumption variants could be intractable in practice. Exploring different or even adaptive methods to select these assumptions could be a possible direction for future work.

Some components in our framework are manually designed, such as the number of agent perspectives or market contexts, which might not capture the full range of heterogeneity present in real-world markets. However, our proposed framework is modular and extensible: additional agents or contexts could be incorporated without modifying any of the core architecture. 
Future work could also explore higher-level coordination mechanisms that adaptively decide the composition of agents and contexts. 
% Notably, this would increase the levels of agent hierarchy in this framework design, and the complexity and computational efficiency would have to be considered.

\end{document}